\documentclass{ar2e_preprint}
\usepackage{ARAstroBib}
\def\lesssim{\mathrel{\hbox{\rlap{\hbox{\lower4pt\hbox{$\sim$}}}\hbox{$<$}}}}
\def\gtrsim{\mathrel{\hbox{\rlap{\hbox{\lower4pt\hbox{$\sim$}}}\hbox{$>$}}}}
\def\micron{\mbox{$\mu$m}}
\def\arcdeg{\mbox{$^\circ$}}

\begin{document}

\input epsf.def   

\input psfig.sty

\def\nat{{\em Nature }}
\def\aap{{\em Astron. \& Astrophys. }}
\def\aj{{\em Astron.~J. }}
\def\apj{{\em Astrophys.~J. }}
\def\araa{{\em Ann. Rev. Astron. Astrophys. }}
\def\apjl{{\em Astrophys.~J.~Letters }}
\def\apjs{{\em Astrophys.~J.~Suppl. }}
\def\icarus{{\em Icarus }}
\def\mnras{{\em MNRAS }}
\def\pasp{{\em Pub. Astron. Soc. Pacific }}

\jname{Annual Review of Astronomy \& Astrophysics}
\jyear{2012}
\jvol{50}
\ARinfo{1056-8700/97/0610-00}

\title{The Formation and Early Evolution of Low-mass Stars and Brown Dwarfs}

\markboth{Kevin L. Luhman}{Formation \& Early Evolution at Low Masses}

\author{Kevin L. Luhman$^{1,2}$
\affiliation{
$^1$ Department of Astronomy and Astrophysics, The Pennsylvania
State University, University Park, PA 16802, USA; kluhman@astro.psu.edu\\
$^2$ Center for Exoplanets and Habitable Worlds, The Pennsylvania State
University, University Park, PA 16802, USA}}


\begin{abstract}

The discovery of large numbers of young low-mass stars and brown dwarfs over
the last decade has made it possible to investigate star formation
and early evolution in a previously unexplored mass regime.
In this review, we begin by describing surveys for low-mass members of
nearby associations, open clusters, star-forming regions and the
methods used to characterize their stellar properties.
We then use observations of these populations to test theories of star
formation and evolution at low masses.
For comparison to the formation models, we consider the initial mass function,
stellar multiplicity, circumstellar disks, protostellar characteristics, and
kinematic and spatial distributions at birth for low-mass stars and
brown dwarfs.
To test the evolutionary models, we focus on measurements of dynamical masses
and empirical Hertzsprung-Russell diagrams for young brown dwarfs and
planetary companions.

Posted with permission from the Annual Review of Astronomy and Astrophysics, 
Volume 50 \copyright 2012 by Annual Reviews, http:// www.annualreviews.org.

\end{abstract}

\maketitle

\section{INTRODUCTION}

Molecular clouds give birth to stars across a wide range of masses.
The most massive stars are born on the main sequence while low-mass stars
must contract for tens to hundreds of millions of years
before becoming hot enough for sustained hydrogen fusion. 
A great deal of observational and theoretical effort has been invested in
understanding star formation and pre-main-sequence evolution
and their dependence on stellar mass.
Over the last decade, it has become possible to extend these studies
to the least massive stars and brown dwarfs as surveys have uncovered them
in large numbers and at a variety of ages.
In this way, star formation and early evolution can be investigated across more
than four orders of magnitude in stellar mass, providing more stringent
tests of the theories for these processes.

In this review, we summarize the theoretical and observational
work on the formation and early evolution of low-mass stars and brown dwarfs.
We define a ``low-mass star" as having a 
mass between $\sim0.2$~$M_\odot$ and the hydrogen burning mass limit
\citep[$\sim0.075$~$M_\odot$,][]{bur97,cha00c}.
All free-floating objects (as well as some companions) below this mass
range are considered brown dwarfs.
Some studies have adopted the deuterium burning limit
\citep[$\sim0.012$~$M_\odot$,][]{cha00a,spi11} as the lower limit for
the definition of a brown dwarf \citep{bas00},
but this is not done here since deuterium burning has a negligible
impact on stellar structure and evolution \citep{cha07ppv}.
Since we are examining the formation and early evolution of low-mass stars and
brown dwarfs, we focus on objects with ages $\lesssim100$~Myr, which is the 
time-scale for low-mass stars to approach the main sequence.
However, we also consider the properties of older stars and brown dwarfs
that can help constrain the formation and evolutionary theories, such as
their multiplicity and mass function.

This article complements previous reviews concerning low-mass stars and
brown dwarfs.
\citet{bas00} reviewed early discoveries of brown dwarfs and \citet{kir05}
reviewed the spectral classification of (mostly older) L and T dwarfs, 
whereas we focus on the discovery and characterization of
low-mass stars and brown dwarfs at young ages.
\citet{whi07} and \citet{cha00c} reviewed theories for the formation
and evolution of brown dwarfs, respectively, and
\citet{luh07ppv} and \citet{bur07ppv} compared the predictions of the
formation models to observations. We summarize the latest developments
in those theories and update the previous reviews of the observational
constraints on the formation of brown dwarfs.

\section{SURVEYS FOR YOUNG LOW-MASS STARS AND BROWN DWARFS}

\subsection{Search Strategies}

Studies of the formation and early evolution of low-mass stars and brown
dwarfs have been enabled by surveys that discover and characterize these
objects, particularly at young ages.
The targets of these surveys have consisted of the nearest young associations,
open clusters, and star-forming regions ($d=50$--500~pc, $\tau=1$--100~Myr).
Young brown dwarfs are brightest at near-infrared (IR) wavelengths
(1--2~\micron, $YJHK$), but because of their relatively warm temperatures, they
often can be detected in red optical bands as well (0.6-1~\micron, $RIZ$).
In fact, an optical filter is quite helpful for identifying red, late-type
objects when combined with data at longer wavelengths.
As a result, surveys for young brown dwarfs often have employed both optical
and near-IR images.
These data have been collected through both dedicated imaging and wide-field
surveys. Examples of the latter include the
Two Micron All-Sky Survey \citep[2MASS,][]{skr06},
the Deep Near-Infrared Survey of the Southern Sky \citep[DENIS,][]{ep99}
the Sloan Digital Sky Survey \cite[SDSS,][]{yor00}, and the United Kingdom
Infrared Telescope Infrared Deep Sky Survey \citep[UKIDSS,][]{law07}.

To identify candidate low-mass stars and brown dwarfs, images
are searched for objects that have colors/magnitudes or proper motions
that are indicative of membership in the target cluster or association.
If the images are deep enough and an appropriate combination of
filters is present, it is actually easier to identify cool dwarfs
than many other classes of astronomical sources because of the distinctive
nature of their spectral energy distributions (SEDs), which results from strong
molecular absorption bands. However, a few words of caution are warranted
regarding photometric searches.
First, some studies have designed their photometric criteria for the
selection of candidates based on theoretical magnitudes and colors
that are predicted for the age of the target cluster or association.
However, model colors are often too inaccurate for this purpose.
Instead, it is better to use the photometry of known young low-mass
objects, either in the target population or in others with a similar age,
to guide the identification of candidates.
If this is not possible, then the theoretical predictions for temperature
and luminosity should be converted to magnitudes and colors using 
empirical relations between spectral types, temperatures, colors,
and bolometric corrections. These relations are incomplete for young objects,
but it is likely better to use the relations for dwarfs rather than rely on 
synthetic magnitudes.
In addition to applying properly designed color and magnitude criteria,
it is important to carefully account for photometric errors during the
selection of candidates. For instance, near the detection limit of
a given set of images, one can find sources that appear to satisfy almost
any color criteria because of the large errors.
A similar problem can arise when candidate low-mass members of clusters are 
identified through poorly constrained proper motions.

In general, it is necessary to obtain spectra of young low-mass candidates
selected from photometry or proper motions to confirm their membership via
measurements of radial velocities or (more commonly) signatures
of youth. Spectra are also important for providing spectral types, which
constrain the temperature and mass of a candidate.
Low-resolution spectra are normally preferred for the initial followup
of candidates since they are adequate for measuring the broad molecular
absorption bands of late-type objects while also providing the highest
signal-to-noise ratios.
For instance, SpeX at the NASA Infrared Telescope Facility \citep{ray03}
offers an ideal configuration of this kind, producing spectra 
from 0.8--2.5~\micron\ with a resolution of $\sim100$.
It is likely that SpeX has observed more brown dwarfs,
both young and old, than any other spectrograph.
Adam Burgasser maintains a compilation of a large number of these spectra
of brown dwarfs in the SpeX Prism Spectral Libraries at
http://www.browndwarfs.org/spexprism.
Searches for brown dwarfs and various other kinds of rare sources (e.g., high
redshift quasars, gamma ray bursts) would benefit greatly if a comparable
observing mode was available on more telescopes. Additional details regarding
the spectroscopic confirmation of young low-mass stars and brown dwarfs are
discussed in Section~\ref{sec:char}.

\subsection{Young Associations and Moving Groups}

The nearest samples of young stars to the Sun ($d<100$~pc) are found in
small moving groups and associations \citep{zuc04,zuc11}.
These stars are no longer associated with molecular clouds, and hence
have ages of $\gtrsim5$~Myr. 
Given their proximity and youth,
they represent the best available targets for surveys to detect
and resolve substellar companions and circumstellar disks.
In addition, since their spectra are relatively unreddened ($A_V<1$),
members of the nearest associations are attractive candidates
for spectral classification standards for young stars and brown dwarfs
\citep{cru09}.

If nearby groups and associations have initial mass functions (IMFs) similar
to those measured in the field, open clusters, and star-forming regions, then
they should contain significant populations of undiscovered low-mass stars
and brown dwarfs based on the numbers of known members at higher masses. 
Surveys for these low-mass members have primarily employed one of two
strategies: searching for candidate members of nearby associations based on
data from wide-field photometric and astrometric surveys, or
checking late-type dwarfs found during surveys of the field for
evidence of youth or membership in any of the nearby groups.
In either strategy, to firmly establish membership of a given candidate,
one seeks to show that it exhibits spectroscopic indicators of youth and
that it shares the same 3-D space motion as the known members,
which requires measurements of parallax, proper motion, and radial velocity.

In the first of the two survey methods mentioned above, candidate
low-mass members of nearby associations are initially identified based
on signatures of youth and activity (X-ray, UV), late spectral type
(optical, near-IR), or kinematic membership (proper motions, radial velocities).
In the discovery of the first known brown dwarfs in a nearby association,
\citet{giz02} used photometry from 2MASS and the Palomar
Observatory Sky Survey to identify candidate late-M and L dwarfs
in the direction of the TW Hya Association (TWA, $\tau\sim10$~Myr).
Through followup spectroscopy, he found that two of the candidates,
2MASSW J1207334-393254 (hereafter 2M~1207-3932) and 2MASSW J113951-3159211,
exhibited spectral features that indicated low surface gravities, and hence
young ages. \citet{moh03b} and \citet{sch05} confirmed the membership of these
objects via measurements of radial velocities and proper motions, respectively.
\citet{sch05} also used their proper motion data to uncover an additional
substellar member of TWA, SSSPM J1102-3431.
These three objects have spectral types near M8.5
and are among the most widely-studied young brown dwarfs, especially
2M~1207-3932 (e.g., Sections~\ref{sec:sptype}, \ref{sec:hr}).
Using photometry and proper motions from wide-field optical and near-IR surveys,
\citet{loo07} discovered a candidate member of TWA with an even later
spectral type of M9.5, DENIS J124514.1-442907.
Its membership seems probable based on the evidence
of youth in its optical and near-IR spectra, although radial velocity and
parallax measurements are needed for a definitive assessment of membership.
With the increasing availability of sensitive astrometric surveys in recent
years, it has become common to use both photometry and proper motions for the
selection of candidate low-mass members of young associations, which are 
then confirmed by radial velocity measurements \citep{cla10,gal10,schl10}.

Although they are not sensitive to members at substellar masses,
X-ray and UV satellites have helped to identify
possible low-mass stars in nearby associations.
X-ray surveys have been performed across a wide range of field sizes.
For instance, deep X-ray images of small fields surrounding the 
B stars $\eta$~Cha and $\epsilon$~Cha have uncovered compact groups of
associated young low-mass stars \citep{mam99,fei03}. Subsequent optical
and near-IR imaging has revealed additional members of these
associations at lower masses and across wider areas
\citep{law02,lyo04,son04,luh04eta1,luh04eta2,mur10}.
Meanwhile, \citet{ria06b}, \citet{loo10}, and \citet{shk09} have used
data from the {\it R\"ontgen Satellite} (ROSAT) All-Sky Survey to identify
candidate young low-mass stars across much larger areas of sky.
It has recently become possible to extend surveys for young stars to
UV wavelengths with the availability of wide-field data from the
{\it Galaxy Evolution Explorer} \citep{mar05}.
\citet{shk11} and \citet{rod11} have found that it seems to offer
better sensitivity to young low-mass stars than the all-sky data from ROSAT.

As noted above, a second approach to finding low-mass members
of nearby associations is to search for them among the cool dwarfs
found in surveys of the field.
Within the hundreds of late-type dwarfs that they have discovered with 2MASS
and other surveys, \citet{kir06,kir08}, \citet{cru07}, and \citet{cru09} have
identified more than two dozen late-M and L dwarfs that have gravity-sensitive
spectral features indicative of youth ($\tau\lesssim100$~Myr), and hence
are promising candidates for substellar members of nearby moving groups.
Similar objects have been found during other spectroscopic surveys of 
late-type field dwarfs \citep{nak04,all10,rei10a}.
These sources have not been definitively established as members of specific
groups since complete sets of parallaxes, proper motions, and radial
velocities are currently unavailable. However, \citet{ric10b} have
shown that one of the young M dwarfs from \citet{kir08}, 2MASS 06085283-2753583,
is a likely member of the $\beta$~Pic moving group ($\tau\sim12$~Myr)
based on the space motion derived from its proper motion, radial velocity,
and spectroscopic distance. \citet{rei09b} has presented similar evidence
suggesting that 2MASS J0041353-562112 (M7.5) may be a member of the
Tucana/Horologium association ($\tau\sim30$~Myr).
Instead of spectroscopic signatures of youth, other studies have
used proper motions \citep{ban07} and radial velocities
\citep{zap07a,sei10} for the initial identification of possible association
members among field L and T dwarfs, but 
neither definitive kinematic evidence of membership nor
confirmation of low gravity via spectroscopy has been presented for these
candidates to date.

\subsection{Young Open Clusters}
\label{sec:open}

Young open clusters span the same range of ages as
the nearest associations and moving groups ($\tau>5$~Myr).
The nearest examples of the former are more distant than associations like
TWA ($d>100$~pc), but they are much richer (100's--1000's of members), providing
better constraints on the statistical properties of stellar populations, such
as their IMF and multiplicity. In addition, open clusters are more
compact on the sky, so they can be imaged in their entirety down to
very faint levels.

The methods of searching for low-mass members of open clusters are similar
to some of those applied to nearby associations and moving groups.
The initial identification of candidate members is usually based on proper
motions and/or optical color-magnitude diagrams.
In early surveys, these measurements were made with photographic plates,
which could detect low-mass stars in the nearest clusters.
The extension of these surveys into the substellar regime became feasible
in the late 1980's with the development of sensitive, large-format CCD cameras.
These devices and their near-IR counterparts are now capable of imaging an
entire cluster ($>10$~deg$^{2}$) in a reasonable amount of time.
Proper motion surveys for substellar members are growing
more powerful over time as the baselines from the first deep
wide-field surveys steadily lengthen \citep{bih06,cas07}.
As in young associations, candidate members of a given open cluster
are confirmed by a combination of proper motions, radial
velocities, and spectral signatures of youth.
A subset of these diagnostics of membership is often sufficient since
open clusters do not overlap with other young populations on the sky
as much as nearby associations, which are much more widely distributed.
This is convenient since radial velocity measurements are not feasible
for the faintest brown dwarf candidates. For these objects, a spectrum
showing features indicative of a cool, low gravity atmosphere is
normally sufficiently convincing evidence of membership.

The most promising open clusters for detecting low-mass stars and
brown dwarfs are those that are youngest and nearest
($\tau<1$~Gyr, $d<200$~pc), which include IC~2391 (50 Myr),
$\alpha$~Per (90 Myr), the Pleiades (100--125 Myr), the Hyades (625 Myr),
and Praesepe ($\sim600$ Myr).
Because it offers the best combination of age, proximity, and richness,
the Pleiades was the first open cluster surveyed for brown dwarfs.
Proper motions measured from photographic plates were used to identify members
down to $\sim0.1$~$M_\odot$ across a large fraction of the cluster
\citep{ham93}.
The first CCD imaging of the Pleiades covered much smaller areas 
($\sim0.1$~deg$^2$) but reached fainter magnitudes than the photographic
data \citep{stau89,jam89}. One of the candidates from \citet{stau89}, PPl~1,
was eventually confirmed as substellar through the Li test \citep{stau98a},
although the first brown dwarfs in the Pleiades confirmed in this manner
consisted of PPl~15 and Teide~1 \citep{stau94,reb95,reb96,bas96,zap97a}.
The Pleiades has been searched extensively for brown dwarfs since those
initial discoveries \citep[][references therein]{mor03,stau07},
producing a census that extends down to L spectral types and masses
of $\sim0.025$~$M_\odot$ \citep[e.g.,][]{mar98b}.
Candidate T-type members with potential masses of $\sim0.01$~$M_\odot$ 
have been found \citep{cas07,cas11}, but they currently lack spectroscopic
confirmation of their cool nature or youth.
Surveys of the other nearby open clusters mentioned above have been quite
successful as well \citep[e.g.,][]{bar04a,lod05b,bou08}.
In fact, it has been difficult for spectroscopic followup to keep pace with
the large number of reported candidates, which is unfortunate since complete
spectroscopic samples are valuable for addressing a number of issues regarding
the formation and evolution of low-mass objects, as discussed elsewhere in
this review.

\subsection{Star-Forming Regions}
\label{sec:sfr}

The nearest star-forming regions ($\tau\lesssim5$~Myr) are roughly comparable
to the nearest open clusters in terms of proximity and richness.
The stellar populations within the former are more difficult to identify and
study because of the dust extinction from their natal molecular clouds.
However, star-forming regions offer an opportunity to detect brown dwarfs
when their luminosities are highest, witness the earliest stages of 
their formation, and examine how their properties (e.g., IMF, multiplicity)
vary with star-forming conditions.
The properties of embedded clusters have been reviewed by \citet{lad03}
and individual nearby regions have been reviewed in the
{\it Handbook of Star Forming Regions} (B. Reipurth, ed.). 
The discussion in this section includes populations with ages of
5--10~Myr like Upper Sco and some of the clusters in Orion, even
though they are no longer embedded within molecular clouds and thus are 
not experiencing ongoing star formation.

Near the same time that CCDs were first used to search for brown dwarfs in the
Pleiades, early near-IR detector arrays were applied to nearby embedded
clusters, often with the same objective.
(CCDs also were capable of detecting the less obscured brown dwarfs 
in star-forming regions, but this was not yet realized.)
IR surveys of Ophiuchus \citep{gy92}, the Trapezium Cluster \citep{mcc94},
and IC~348 \citep{lad95} detected objects that are now known to be
probable brown dwarfs, but they were not identified as candidates at
that time. This is because near-IR broad-band filters are less sensitive
to spectral type than the optical bands used in open clusters, making
it more difficult to distinguish brown dwarfs from field stars.
The presence of variable reddening towards the cluster members
and background stars in star-forming regions further hampers reliable
selection of candidates. Nevertheless, a few individual brown dwarf
candidates were reported from IR imaging of embedded clusters
\citep{com93,rie90}.
GY141 ($\rho$ Oph 162349.8-242601) was one of the first candidates that
was spectroscopically confirmed as a young, late-type object ($>$M6).
\citet{rie90} suggested that it was a foreground dwarf because its near-IR
colors did not exhibit any reddening, but \citet{com98} identified it
as a possible young brown dwarf based on the detection of mid-IR excess emission
at 4.5~\micron\ with the {\it Infrared Space Observatory} ({\it ISO}),
which indicated the presence of a circumstellar disk. \citet{luh97} used optical
spectroscopy to confirm its youth and late spectral type (M8.5).
Although it was a small aspect of a much larger study of the Orion Nebula
Cluster (ONC), and hence received little attention at the time, 
\citet{hil97} also presented some of the first spectral classifications 
of young late-type objects.

Rather than search for individual brown dwarf candidates, some of the
early surveys of embedded clusters focused on statistical estimates
of the substellar IMF based on IR luminosity functions.
This approach was applied most extensively to the Trapezium Cluster
at the center of the ONC because it is very compact and is located
in front of a dense molecular cloud, reducing the contamination from
background stars \citep{luh00trap,hil00,luc00,mue02,luc05}.
Other clusters have been studied in this way as well \citep{com93,mue03},
but because they lack the uniquely favorable geometry of the Trapezium,
background stars dominate the observed luminosity functions at the magnitudes
of brown dwarfs, and hence the resulting constraints on the substellar
IMF have large uncertainties.

To find promising candidates for low-mass stars and brown dwarfs in
star-forming regions, several methods have been employed over the last decade.
Because of the extreme youth of these objects, they 
can be identified via signatures of activity (X-rays),
accretion (H$\alpha$, UV), and disks (mid-IR).
Deep X-ray images are capable of detecting most of the low-mass stars
in the nearest embedded clusters, but they reach few of 
the brown dwarfs \citep{fei04,stel04,sce07}.
Although H$\alpha$ imaging has not been used widely for this purpose,
it did uncover some of the first spectroscopically-confirmed late-type
members of star-forming regions \citep{com99}.
As described earlier, mid-IR excess emission was used to identify GY141
as a possible disk-bearing brown dwarf. 
The ability of the {\it Spitzer Space Telescope} \citep{wer04}
to collect sensitive, wide-field mid-IR images
has made it possible to apply this technique on a much larger scale
\citep{luh06tau2,luh09tau,mue07,all07,reb10}.
Data from {\it Spitzer} have been particularly crucial for detecting 
heavily obscured, protostellar brown dwarfs (see Section~\ref{sec:proto}).

\begin{figure}
\centerline{\psfig{figure=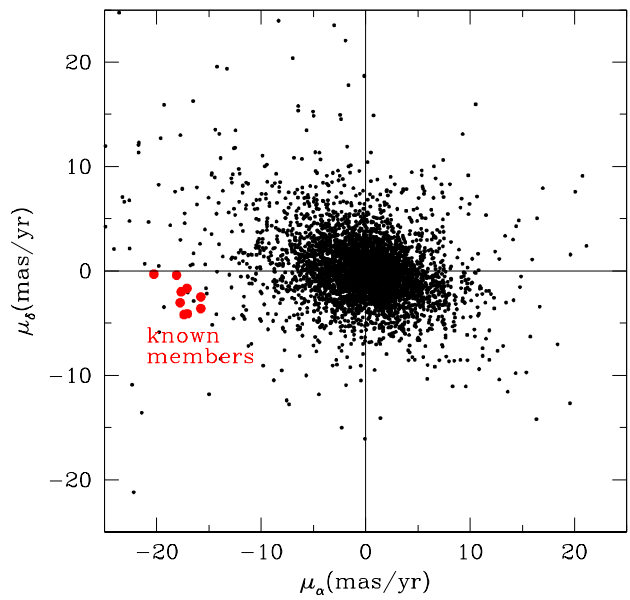,height=40pc}}
\caption{Proper motions of stars in Chamaeleon~I measured from two epochs
of {\it HST}/ACS images (J. Bochanski, in preparation). The motions of the
known members within these images are indicated ({\it large red points}).}
\label{fig:pm}
\end{figure}

As in open clusters, candidate low-mass members of star-forming regions
also can be identified with proper motions and color-magnitude diagrams 
measured from optical and near-IR images. These data are generally
more sensitive and unbiased than the surveys based on youth indicators.
Proper motion selection of candidates has not been widely used
in star-forming regions since extinction prevents
many of the members from appearing in photographic plates.
However, the earliest wide-field images with CCDs and near-IR detectors
are growing old enough that sensitive proper motion surveys for low-mass
members are becoming feasible. Such proper motion
measurements are especially accurate if one or more epochs of images
have been obtained with the {\it Hubble Space Telescope} ({\it HST}) given its
high spatial resolution. Figure~\ref{fig:pm} shows an example of how
{\it HST} can be used for this purpose.
Ideally, if one could map the nearest star-forming regions with the
resolution of {\it HST} and the wavelength coverage and sensitivity of
{\it Spitzer} (e.g., {\it James Webb Space Telescope}) across a period of
several years, virtually all of the members could be identified through
proper motions.

\begin{figure}
\centerline{\psfig{figure=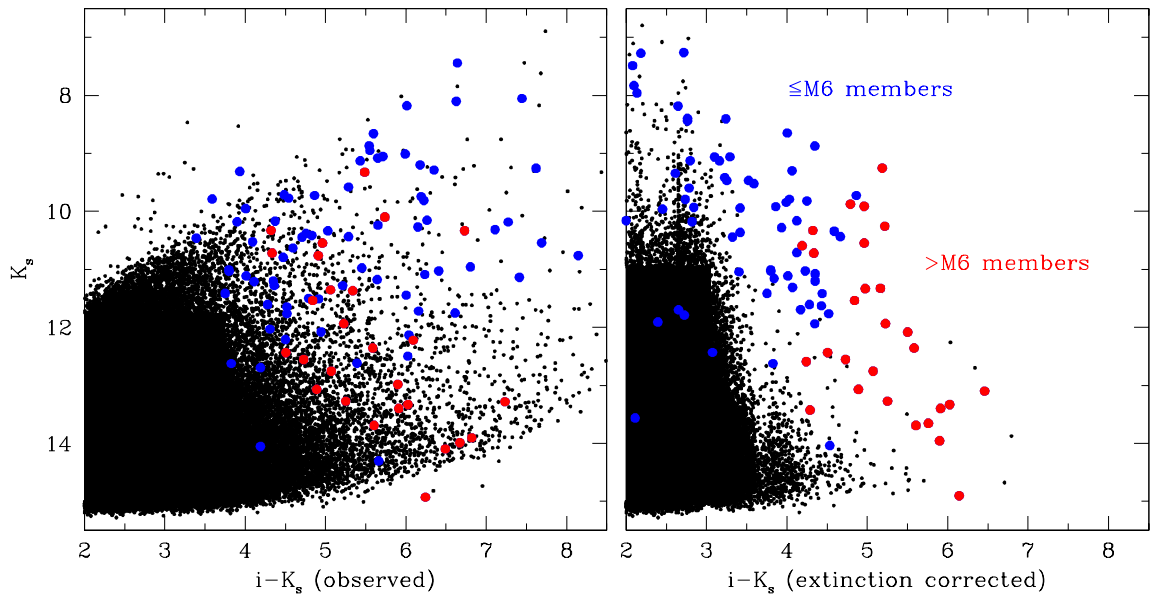,height=40pc}}
\caption{Observed and extinction-corrected color-magnitude diagrams
for the portion of Taurus imaged by SDSS \citep{fin04}. The known
members of Taurus are indicated ({\it large blue and red points}).
The $i$ and $K_s$ data are from SDSS and 2MASS, respectively.
}
\label{fig:cmd}
\end{figure}

Color-magnitude diagrams have proved to be the most successful tool for finding 
low-mass members of star-forming regions.
Since clusters with ages of $>3$~Myr usually 
have little extinction ($A_V<2$), standard color-magnitude diagrams like those
used for open clusters are sufficient for the selection of candidates.
Examples of such populations include $\sigma$~Ori \citep{bej99},
$\lambda$~Ori \citep{bar04b}, and Upper Sco \citep{sle06}.
However, extinction from the parent molecular cloud in younger regions can be
quite large, and it varies significantly from one line of sight to another
($A_V=0$--50), which increases the contamination of reddened background stars
in the area of a color-magnitude diagram inhabited by cluster members.
Late-type members can be distinguished from these background stars 
by inspection of a color-color diagram like $I-K$ versus $J-H$
in which the reddening vector is roughly perpendicular
to the sequence of dwarf colors \citep{luh00tau}.
To represent the membership constraints from a color-magnitude diagram
and $I-K$ versus $J-H$ more compactly, extinctions of individual stars
can be estimated from the latter, which are then used to deredden the
locations of stars in the color-magnitude diagram \citep{luh03tau}.
The dramatic reduction in contamination from background stars is
illustrated by the comparison of observed and extinction-corrected
color-magnitude diagrams for Taurus in Figure~\ref{fig:cmd}. 
Thus, the most refined selection of candidate low-mass stars and brown dwarfs
is produced when data at both optical and near-IR bands are available.
2MASS was an essential source of the IR data in early surveys, particularly in
the larger regions like Taurus \citep{bri02}. UKIDSS is now providing
much deeper IR images for several nearby open and star-forming clusters
\citep{lod06}.
Although the members of the most embedded clusters like Ophiuchus are
often below the detection limits of optical images, it has been possible to
identify candidate substellar members with $JHK$ photometry alone \citep{alv10}.
Specialized filters that are designed to measure the near-IR absorption
bands from H$_2$O and CH$_4$ also have been used to search for cool
members of star-forming regions \citep{burg09,hai10,pen11}.
As in open clusters, candidates are usually confirmed through low-resolution
optical or near-IR spectroscopy that demonstrates a late spectral type
and a young age.

The nearest and richest stellar populations with ages of $\lesssim10$~Myr
are found in Taurus, Perseus (IC~348, NGC~1333), Chamaeleon, Orion
(ONC, $\sigma$~Ori, $\lambda$~Ori), Ophiuchus, Lupus,
and Upper Sco.  Surveys of these regions have discovered a few hundred
members that are likely to be brown dwarfs ($>$M6) with spectral
types as late as early L \citep{zap00,luh08cha1,luh09tau,lod08,wei09}.
Candidate young T dwarfs also have been found in $\sigma$~Ori
\citep{zap02b,zap08} and Ophiuchus \citep{marsh10a}.
The membership of the former object is uncertain \citep{bur04} while the
source in Ophiuchus is too reddened for a foreground dwarf, seems too bright
for a background dwarf, and does not match the spectra of normal T dwarfs,
indicating that it may be a young member of the cloud.
The samples of spectroscopically confirmed young low-mass stars and
brown dwarfs in Taurus and Chamaeleon are particularly valuable because of
their proximity, relatively low extinction, large sizes, lack of crowding and
bright nebular emission, and the uniformity of their spectral classifications
\citep[][references therein]{luh08blue,luh10tau1}.
Upper Sco is also very promising; based on the number of known members above
$\sim0.5$~$M_\odot$, it should contain the largest substellar population
of any star-forming region within 150~pc ($>200$ brown dwarfs).
Deep, wide-field surveys of Upper Sco through UKIDSS are beginning
to realize its potential \citep{lod06,lod08,lod11}.

\subsection{Characterization of Stellar Properties}
\label{sec:char}

After a young low-mass star or brown dwarf is discovered,
we can begin to study it in detail by estimating its basic stellar
properties, such as its spectral type, effective temperature,
extinction, bolometric luminosity, mass, and age. The methods for doing so
are reviewed in this section. An extensive review of the characterization
of older, cooler brown dwarfs can be found in \citet{kir05}.

\subsubsection{SPECTRAL TYPE}
\label{sec:sptype}

The measurement of a spectral type is important for confirming that
a given candidate is cool and young, and is not a field star or a galaxy.
In addition, because low-mass stars and brown dwarfs are expected to
maintain roughly constant temperatures during their early evolution,
spectral types should comprise good observational proxies of stellar masses.

At ages of $\lesssim100$~Myr, objects at the hydrogen burning mass limit
are predicted to have temperatures near 3000~K \citep{bur97,bar98},
corresponding to a spectral type of $\sim$M6. This prediction appears to be
consistent with measurements of dynamical masses and Li abundances 
\citep[][Sections~\ref{sec:dyn}]{bas96,bas99,stas06,clo07b}.
Young brown dwarfs should have progressively later types with decreasing
mass, and could extend into the L and T spectral classes.
For instance, evolutionary models predict that a brown dwarf
with a mass of 2~$M_{\rm Jup}$ should have a temperature of $\sim1200$~K
at an age of 1~Myr, which corresponds to a spectral type of early T for
field dwarfs \citep{kir05}. However, the onset of methane absorption
(i.e., a type of T0) may occur at lower temperatures for young brown dwarfs 
than for older objects in the field \citep[][Section~\ref{sec:hr}]{bar11b}.

The spectral types of field M and L dwarfs are defined by the strengths
of various atomic and molecular absorption features (e.g., TiO, VO)
that appear at red optical wavelengths \citep{hen94,kir91,kir97,kir99a,kir05}.
Averages of optical spectra for dwarf and giant standards match well with the
spectra of young late-M objects, and thus have been used to define the
spectral types of these sources \citep{luh97,luh99ic}.
Since L-type giants do not exist, a preliminary classification system
for young L dwarfs has been defined at optical wavelengths based
on a comparison to normal L dwarfs \citep{cru09}.
The standards for this scheme consist of nearby field dwarfs that have been
identified as young based on their gravity sensitive lines
\citep{cru07,cru09,kir06,kir08}. L-type members of young clusters generally
are less attractive as standards since they are fainter and can have
significant reddening, although they do offer the advantage of known ages
via their cluster membership.
For field T dwarfs, the primary classification scheme is based on near-IR
H$_2$O and C$_2$H$_4$ bands \citep{bur06}. The normal T dwarf standards
are well-suited for classifying the moderately young T dwarf HN Peg B
\citep[$\tau\sim300$~Myr,][]{luh07tdwarfs}.
The classification of younger T dwarfs has not been explored since
only a few candidates have been identified \citep{zap02b,marsh10a}.

\begin{figure}
\centerline{\psfig{figure=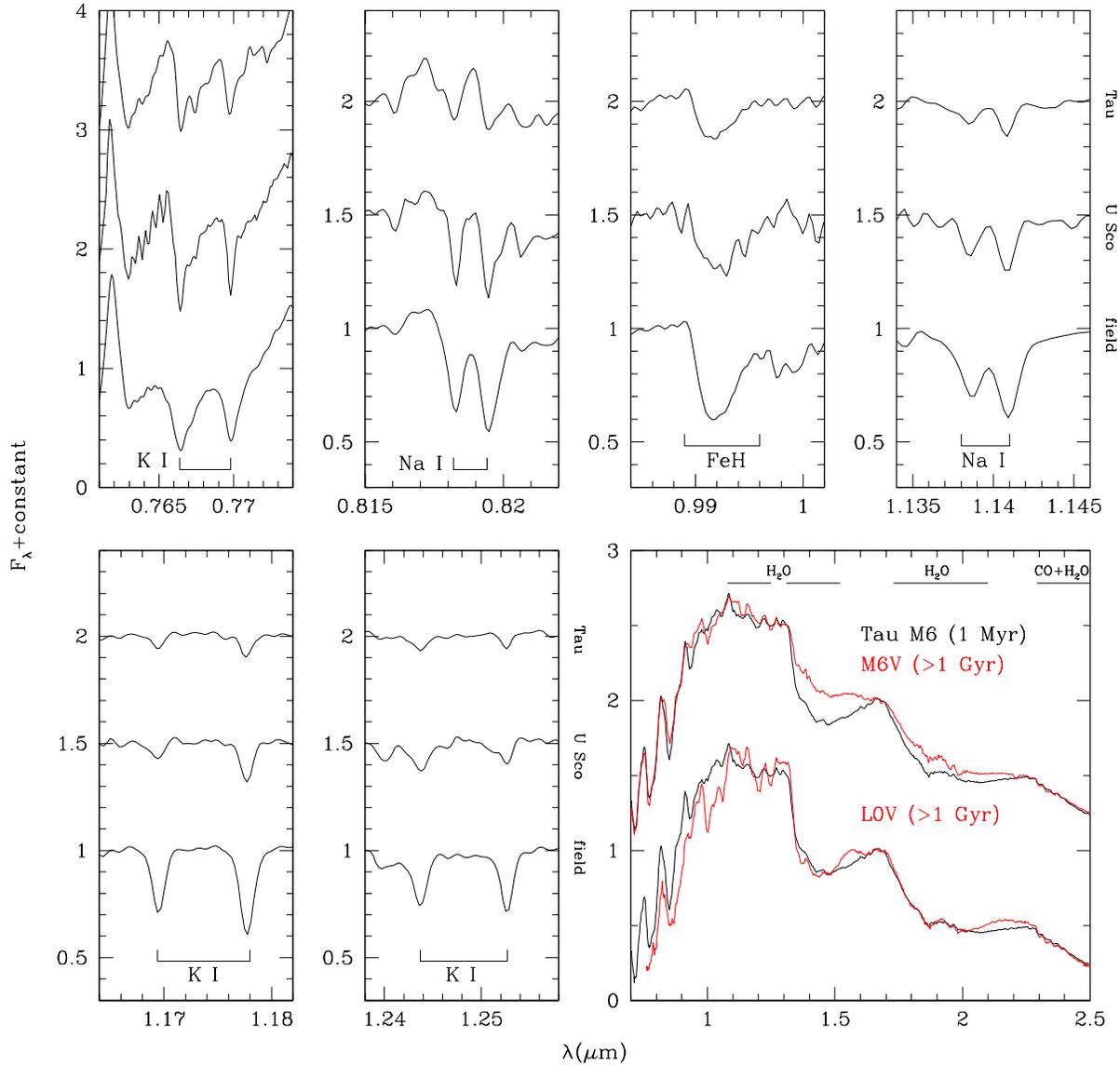,height=40pc}}
\caption{
The top and left panels show gravity-sensitive absorption lines 
for late-M members of Taurus ($\tau\sim1$~Myr), Upper Sco ($\tau\sim12$~Myr),
and the field \citep[$\tau\gtrsim1$~Gyr,][]{luh07oph}. In the lower right
panel, a low-resolution spectrum of an M6 member of Taurus is compared
to data for M6V and L0V field dwarfs, illustrating how the strength
of H$_2$O absorption and the shape of the $H$- and $K$-band continua
depend on gravity.
}
\label{fig:ir1}
\end{figure}

Although their spectral types have been defined according to optical features,
young late-M and L dwarfs are more easily observed at near-IR wavelengths,
where they are brightest.
To measure IR spectral types that are tied to 
the optical classification system, they should be based on
1) a comparison to dwarf standards for spectral features that do not depend
on gravity or 2) a comparison to optically-classified sources with roughly the
same age for features that depend on both gravity and temperature
\citep[e.g.,][]{luh08blue,luh09tau,cru09}.
Since the H$_2$O bands are the most prominent features in low-resolution
near-IR spectra at M and L types, and since they do vary with both gravity and
temperature, the latter approach is usually necessary.
To illustrate this point, Figure~\ref{fig:ir1} compares the spectra
of an M6 member of Taurus, V410~X-ray~3, and a standard M6V field dwarf, Gl~406.
The M6V dwarf matches the optical absorption bands but
has weaker H$_2$O absorption.
A much cooler field dwarf near L0V is required to reproduce the depth of
H$_2$O in the low-gravity M6 object, as shown in Figure~\ref{fig:ir1}.
Thus, using dwarfs as the standards for classifying young objects via
their H$_2$O bands produces spectral types that are too late \citep{luh03ic}.

Gravity-sensitive spectral features like H$_2$O complicate the
measurement of spectral types, but they also provide valuable constraints
on the ages of late-type dwarfs. In particular, these features offer
a means of confirming the youth, and hence the membership, of candidate
low-mass members of young clusters and associations.
Early surveys for young brown dwarfs used optical spectroscopy of
Na~I, K~I, CaH, and VO for this purpose \citep{ste95,mar96,luh97},
and similar features were soon recognized at near-IR
wavelengths as well \citep{luh98vx,gor03}.
The variation of gravity-sensitive lines with age has been
examined in detail in several studies \citep{mcg04,kir06,ric10a},
which have shown that differences of $\sim1$~dex in log(age) can be
detected within a sample of objects \citep{kir08}.
This is illustrated in Figure~\ref{fig:ir1}, which compares transitions from 
Na~I, K~I, and FeH for late-M members of Taurus ($\tau\sim1$~Myr),
Upper Sco \citep[$\tau\sim12$~Myr][]{pec11}, and the field ($\tau\gtrsim1$~Gyr).
Spectral features of this kind have been used to define gravity classes
that are denoted by the suffixes $\alpha$, $\beta$, $\gamma$, and $\delta$
(e.g., L0$\delta$), which correspond roughly to ages of $\gtrsim1$~Gyr,
$\sim100$~Myr, $\sim10$~Myr, and $\sim1$~Myr, respectively
\citep{kir05,kir06,cru09}.
Figure~\ref{fig:ir1} also shows how the shape of the $H$-band
continuum varies with gravity, exhibiting a broad plateau in old dwarfs
and a triangular peak in young objects \citep{luc01}.
Because it can be detected in low-resolution spectra, this is a particularly
useful gravity diagnostic.

\begin{figure}
\centerline{\psfig{figure=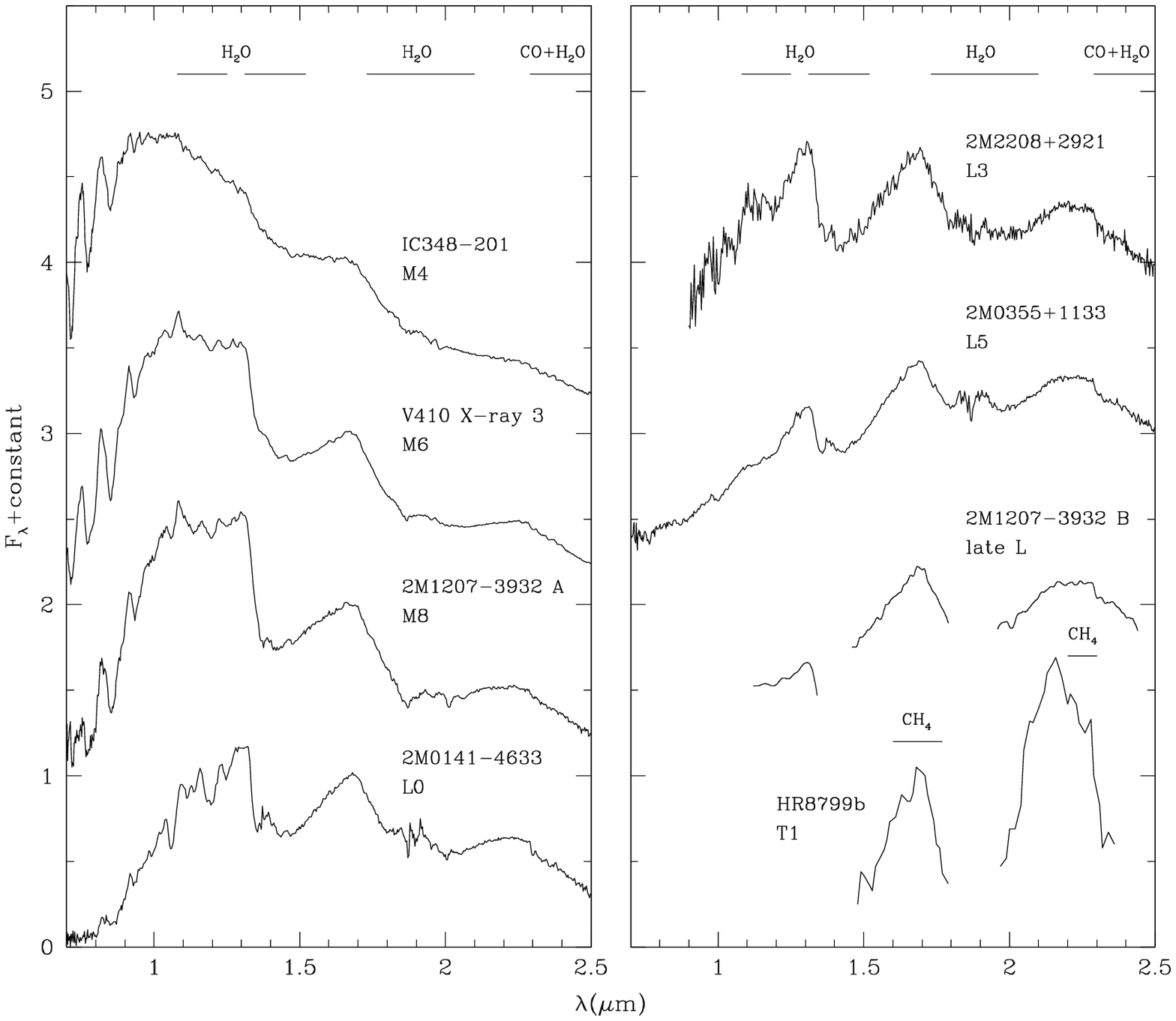,height=40pc}}
\caption{
Low-resolution spectra of young objects ($\tau\sim1$--50~Myr)
from mid-M to early T \citep{kir06,mue07,mor08,pat10,bar11a}.
The data are normalized at 1.68~\micron.
}
\label{fig:ir2}
\end{figure}

To illustrate how the optical and near-IR SEDs
of young low-mass objects vary with spectral type, Figure~\ref{fig:ir2}
shows low-resolution spectra of young sources from mid-M to early T.
This sample consists of members of IC~348 \citep[source 201,][]{luh99ic},
Taurus \citep[V410 X-ray 3,][]{luh98vx}, and TWA
\citep[2M~1207-3932 A and B,][]{giz02,chau04}, three young
field dwarfs \citep[2M~0141-4633, 2M~2208+2921,
2M~0355+1133,][]{kir06,cru09}, and the young companion HR~8799~b
\citep[$\tau\sim60$~Myr,][]{mar08}.
The spectral types of the earliest six sources were measured from optical
spectra, but only near-IR spectra are available for 2M~1207-3932~B and
HR~8799~b.
The latter has been classified as early T based on the tentative
detection of weak CH$_4$ absorption in the spectrum in Figure~\ref{fig:ir2}
\citep{bar11a}. Given its strong H$_2$O bands and its lack of CH$_4$,
2M~1207-3932~B is probably an L dwarf, but a more accurate
classification has not been possible previously because it is much
redder from 1--2~\micron\ than normal L dwarf standards \citep{moh07,pat10}.
As discussed at later points in this review, young L dwarfs like those
in Figure~\ref{fig:ir2} are also unusually red, which suggests that
the red color of 2M~1207-3932~B is related to its youth. 
2M~1207-3932~B closely matches the young L5 dwarf in Figure~\ref{fig:ir2}
from 1.4--2.5~\micron, but is redder in the $J$-band, indicating
that it is probably later than L5. Therefore, we adopt a type of L6--L9
for it. The spectral behavior of objects
at very low masses and young ages like 2M~1207-3932~B and HR~8799~b
will be discussed further in Section~\ref{sec:hr}.

\subsubsection{EFFECTIVE TEMPERATURE}
\label{sec:teff}

Effective temperatures of young low-mass objects have been estimated from
spectral types and from fitting of spectral lines with the predictions
of model atmospheres. For the first approach, a conversion between
spectral types and temperatures is required. To develop a temperature
scale that is appropriate for use with one of the most widely-used
sets of evolutionary models \citep{bar98}, \citet{luh99ic} modified the
scale for M dwarfs so that the components of the young quadruple
system GG~Tau \citep[K7/M0.5/M5.5/M7.5,][]{whi99} exhibit the same iscochronal
ages. \citet{luh03ic} then adjusted this scale further so that the sequences
of IC~348 and Taurus at $\leq$M9 were parallel to those model isochrones on the
Hertzsprung-Russell (H-R) diagram. The resulting scale may differ significantly
from the true conversion, but in the absence of a robust determination
of the temperature scale at young ages (e.g., eclipsing binaries), this
scale offers a reasonable means of interpreting spectral types and
luminosities in terms of masses and ages with evolutionary models.
No attempt has been made to extend the temperature scale from
\citet{luh03ic} to young L dwarfs since few of these objects have been
found in young clusters, but it appears that they may have significantly
different temperatures than dwarfs at a given spectral type
(Section~\ref{sec:hr}).

As an alternative to applying a temperature scale to spectral types,
temperatures (and surface gravities) for young low-mass objects have been
estimated by fitting absorption lines in high-resolution spectra
with lines produced by model atmospheres \citep{moh04a,moh04b,ric10a}.
The resulting parameters
are often consistent with the temperatures implied by spectral types and 
the gravities derived from evolutionary models, but their values do
depend on the transitions in question and the adopted models.
In addition, spectroscopy with sufficient resolution for detailed
model fitting is feasible for only the brightest young brown dwarfs.

\subsubsection{EXTINCTION}
\label{sec:av}

Young brown dwarfs in the solar neighborhood have negligible extinction
($A_V<0.5$), but members of star-forming regions are often obscured by their
natal molecular clouds and (less frequently) circumstellar dust. Older stars in
open clusters and associations also can have noticeable extinction ($A_V\sim1$)
if they are beyond the edge of the Local Bubble ($d\gtrsim100$~pc).
Estimating the amount of extinction toward young low-mass stars and
brown dwarfs is necessary for measuring
their bolometric luminosities. Reddening from extinction also must
be considered during spectral classification since it significantly affects
spectral slopes. Indeed, rather than
use the slope to help constrain the spectral type, one typically must use
it to estimate the extinction if significant obscuration is possible.

In early studies, extinctions for young low-mass stars and brown dwarfs were 
estimated by comparing their broad-band colors to the typical values for
field dwarfs at a given spectral type. Samples of young objects 
are now large enough for types earlier than L0 that it has become possible
to estimate their average intrinsic colors as a function of spectral type by
assuming that the bluest sources at each spectral type have $A_V\sim0$
\citep{luh10tau1}. Since they should be unreddened, young M and L dwarfs
in the solar neighborhood also have constrained the intrinsic colors
at young ages \citep{cru09}. Although colors are more widely used for
estimating extinctions, the most accurate measurements are provided by
the slopes of broad-coverage spectra (e.g., Figure~\ref{fig:ir2})
in comparison to unreddened standards.

\subsubsection{BOLOMETRIC LUMINOSITY}
\label{sec:bc}

Many young low-mass stars and brown dwarfs lack photometry 
across a wide enough range of wavelengths for direct calculations of
their bolometric luminosities. As a result, luminosities are
usually estimated by applying bolometric corrections derived for field
dwarfs to broad-band magnitudes, such as $J$, $H$, or $K$ \citep{gol04}.
This approach seemed reasonable since the near-IR colors of young M dwarfs
are similar to those of field M dwarfs \citep{luh99ic}.
However, beyond a spectral type of $\sim$M9, young objects become redder
than their older counterparts in $J-K$
\citep{bri02,kir06,kir08,cru09,fah09,all10,sch10,bih10}.
This trend is evident in a comparison of near-IR spectra of young L dwarfs like
those in Figure~\ref{fig:ir2} to spectra of field dwarfs, and it applies to
the young planetary companions of HR~8799 as well \citep{mar08}, which
probably have spectral types near the L/T transition \citep{bar11a}.
Similarly, the colors between near- and mid-IR bands for young L dwarfs
in the field are unusually red \citep{zap10,zap11}.
The young T dwarf HN~Peg~B also exhibits red near- to mid-IR colors,
although its near-IR colors are normal \citep{luh07tdwarfs}.
Thus, the bolometric corrections for young L and T dwarfs are probably
not the same as those for standard field dwarfs.
As a result, luminosities based on dwarf-like bolometric corrections are
probably underestimated, which may partially explain why young L dwarfs are 
underluminous compared to normal dwarfs in near-IR bands \citep{fah11b} and why
the sequences of young clusters on H-R diagrams are underluminous relative to
model iscohrones at the latest types \citep[][Section~\ref{sec:hr}]{luh08cha1}.
Near- and mid-IR data are now available for a sufficiently large number
of young low-mass objects that one could estimate the typical bolometric
corrections as a function of spectral type at young ages, but this has
been done only for a few specific spectral types to date \citep{tod10,zap10}.

\subsubsection{MASS AND AGE}

Masses and ages for young low-mass stars and brown dwarfs are normally
estimated by comparing their temperatures and luminosities to the values
predicted by theoretical evolutionary models (Section~\ref{sec:hr}).
The estimates for a given object depend significantly on the details of this
process, such as the adopted temperature scale and evolutionary models,
but it is possible to arrive at masses and ages for young M dwarfs
that are roughly consistent with measurements of dynamical masses, the
expected coevality of components of multiple systems, and other
observational constraints \citep{luh06ab}.
However, the errors in these masses and ages are likely rather high at
L and T types where the SEDs are peculiar
and the atmospheric models are more uncertain.
As a result, whenever possible, it is preferable to discuss young low-mass
objects in terms of observational parameters that are likely to be
correlated with mass and age, namely spectral types and gravity-sensitive
spectral lines (Section~\ref{sec:imf}, \ref{sec:hr}).

\section{FORMATION AT LOW MASSES}

\subsection{Theory}
\label{sec:theory}

Several mechanisms have been proposed for the formation of low-mass
stars and brown dwarfs \citep{whi07,bon07}. They can be summarized as follows:

\begin{enumerate}

\item
Gravitational compression and fragmentation of gas in a massive
collapsing core produces fragments over a wide range of masses.
The tidal shear and high velocities within the cluster prevent the low-mass
objects from accreting to stellar masses \citep{bon08}.

\item
Dynamical interactions among fragments or protostars in a massive core
lead to the ejection of some of them from the core, which prematurely
halts their accretion \citep{rei01,bos01,bat02,bat03,goo04,bat05a,umb05,bat09a}.

\item
Instead of ejection, photoionizing radiation from OB stars halts accretion
by removing much of the envelope and disk of a low-mass protostar
\citep{hes96,whi04}.

\item
The gravitational fragmentation of massive circumstellar disks around stars
produces low-mass companions. Some of these objects are ejected
through dynamical interactions with other companions or nearby stars
\citep{bat02,ric03,bat03,bat05a,goo07,whi06,sta07,sta09,sta11b,off08,off09,she10,thi10}.

\item
Turbulent compression and fragmentation of gas in a molecular cloud produces
collapsing cores over a wide range of masses \citep{pad02,pad04,hen08,elm11}.
The mass of each core determines the mass of the resulting star.
Low-mass stars and brown dwarfs arise from the smallest cores.

\end{enumerate}

Observations of low-mass stars and brown dwarfs that can potentially be used 
to test these theories include the shape of the low-mass IMF and its minimum
mass, stellar multiplicity, spatial and kinematic distributions at birth,
the prevalence and sizes of circumstellar disks and envelopes, and the
dependence of these properties on star-forming environment. 
The number and specificity of predictions for comparison to these data
vary considerably among the models. \citet{rei01} and \citet{bat02,bat03}
have led in this respect, providing valuable motivation for a wide
range of observational studies over the last decade.
Unfortunately, when available, some of the predictions for a given
property are rather similar, limiting the ability of observations to
discriminate among models. For instance, nearly all of the newer models predict
that the velocities of newborn stars and brown dwarfs are indistinguishable,
that brown dwarfs have low binary fractions, and that the IMF is roughly
log-normal with a characteristic mass of $\sim0.5$~$M_\odot$.
In addition, although published values for the minimum mass for
opacity-limited fragmentation range from 0.001--0.01~$M_\odot$
\citep{ree76,low76,sil77,bos88,bat02,boy03,bat05a,whi06,whi07},
a given set of models is often consistent with a fairly wide range of values.
The dependence of the minimum mass on star-forming conditions may represent
a better test of the models, but the number statistics available in young
clusters will limit the accuracy of such measurements.
Predictions do differ significantly for the frequency of wide binaries,
the distribution of binary eccentricities, the ability of brown dwarfs to form
in isolation, and the formation and survival of protostellar brown dwarfs.
In the remainder of this section, we review the current observational
constraints on these properties and the others mentioned above, and
compare them to theoretical predictions.

\subsection{Initial Mass Function}
\label{sec:imf}

In this section, we summarize measurements of the IMF of low-mass stars
and brown dwarfs in the solar neighborhood and the nearest young clusters
and discuss the insight they provide into the formation of these objects.
\citet{bas10} also have described some of these studies as a part of a
larger review of evidence for variations in the IMF.

\subsubsection{SOLAR NEIGHBORHOOD AND GALACTIC DISK}
\label{sec:field}

Measurements of the IMF are frequently characterized in terms of the power-law
forms $dN/dM\propto M^{-\alpha}$ or $dN/dlog M\propto M^{-\Gamma}$, where
$\alpha=\Gamma+1$ and the Salpeter value is $\alpha=2.35$.
The mass function of stars in the solar neighborhood and the Galactic disk
has been found to rise with a roughly Salpeter slope down to 
$\sim0.5$~$M_\odot$, where it then flattens by $\Delta\alpha\sim1.5$
\citep{cha01,kro02,rei02,dea08,cov08,boc10}. Similar IMFs have been
measured in young clusters, as discussed in the next section.

Since the photometric properties of field brown dwarfs are distinct from
those of most low-mass stars, their mass function has been estimated through
a separate set of surveys.
Based on L and T dwarfs uncovered by wide-field surveys like 2MASS,
SDSS, and UKIDSS, the substellar mass function in the solar neighborhood
exhibits a slope of $\alpha\sim0$
\citep{rei99,cha02,all05,met08,pin08,burn10,rey10}.
In comparison, slopes reported for young clusters tend to be near
$\alpha\sim0.5$ (Section~\ref{sec:young}).
This difference is probably not significant since the substellar mass
functions in the field and clusters have been derived with different
methods and thus are subject to different systematic errors.
There is no evidence of the surplus of brown dwarfs in the
field compared to young clusters that is expected if brown dwarfs are born
with much higher space velocities than stars \citep{kro03b,mor05}.

The recent all-sky mid-IR imaging survey by the Wide-field Infrared
Survey Explorer \citep[WISE,][]{wri10} offers an opportunity to
further improve measurements of the substellar mass function in the field.
Data from WISE have already been used to discover $\sim100$ late T dwarfs
\citep{kir11} and objects at even cooler temperatures,
known as Y dwarfs \citep[$T_{\rm eff}<500$~K]{cus11}.
Brown dwarfs at $T_{\rm eff}\lesssim300$~K are particularly valuable
since their numbers are sensitive to minimum masses of the IMF
in the range of 5--10~$M_{\rm Jup}$ \citep{bur04b,kir11}.
Only $\sim1$ field brown dwarf of this kind has been found to date,
but the WISE surveys are still ongoing and the statistical constraints
on the minimum of the IMF should improve.

\subsubsection{YOUNG CLUSTERS}
\label{sec:young}

\begin{figure}
\centerline{\psfig{figure=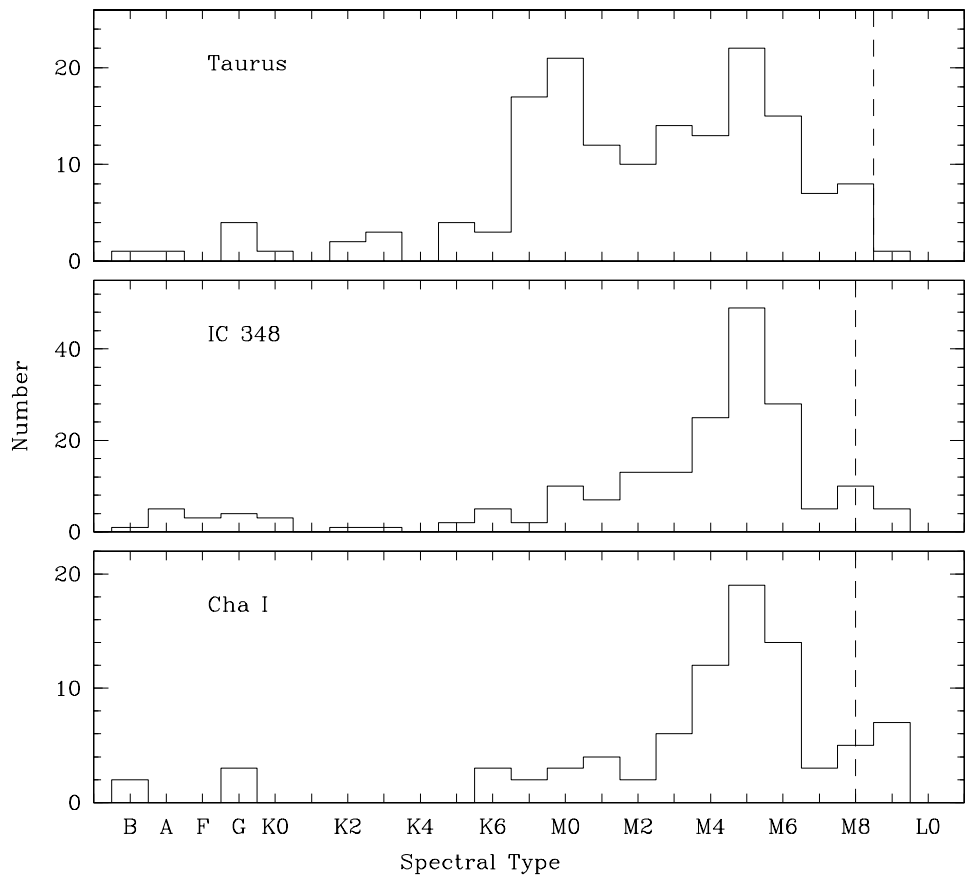,height=40pc}}
\caption{
Distributions of spectral types for representative samples of members of
Taurus ($\tau\sim1$~Myr), IC~348 ($\tau\sim2$--3~Myr), and Chamaeleon~I
\citep[$\tau\sim2$--3~Myr,][]{luh03ic,luh07cha,luh09tau}.
The completeness limits of these samples are indicated ({\it dashed lines}). 
}
\label{fig:histo}
\end{figure}

The surveys for low-mass stars and brown dwarfs in young open clusters
and star-forming regions described in Sections~\ref{sec:open} and \ref{sec:sfr}
have produced a large number of IMF measurements 
\citep[e.g.,][]{bej01,bri02,bar04a,mor04,sle08,lod11}.
Most of the estimated IMFs are fairly similar with $\alpha\sim0.5$ at
$M\lesssim0.2$~$M_\odot$.
There have been a few reports of variations among these data, 
but reliably detecting IMF variations requires that all of the IMFs
subject to comparison have been 1) derived with the same evolutionary
models, spectral classification system, and temperature scale,
2) assessed rigorously for completeness, and 3) measured from large enough
fields to avoid effects of mass segregation.
Claims of variations in low-mass IMFs also should be treated with skepticism
unless they are evident in a direct comparison of the data (e.g., spectral
types) on which the IMFs are based. Given these considerations, 
we find that the only significant variation in published low-mass IMFs
is the unusually high peak mass in Taurus
\citep{bri02,luh03ic}. This variation is illustrated in
Figure~\ref{fig:histo}, which shows a comparison of spectral type
distributions for representative samples of members of Taurus, IC~348,
and Chamaeleon~I
\citep{luh03ic,luh07cha,luh09tau}. The distributions in IC~348 and Chamaeleon~I
peak at M5 while Taurus exhibits a large surplus of K7--M1 stars
($\sim0.8$~$M_\odot$).
It would be quite interesting to measure distributions of spectral types
with the same classification system and comparable completeness for
additional clusters. Because of the narrow widths of the peaks
at M5 in IC~348 and Chamaeleon~I, it should be possible to detect
fairly small variations in the spectral type of this peak (i.e., the IMF's
characteristic mass) when comparing clusters at similar ages.
In addition, a comparison of clusters across a range of ages (e.g., IC~348
vs. the Pleiades) could reveal how spectral type evolves with age for the
mass corresponding to the maximum of the IMF, assuming that the IMF is
invariant within the sample of clusters. Currently, the available spectral
types for low-mass members of most clusters are not sufficiently
numerous, representative, or uniformly measured for these experiments.

The samples from Figure~\ref{fig:histo} for Taurus and Chamaeleon~I
are shown in terms of IMFs in Figure~\ref{fig:imf}, which have been
derived with the evolutionary models of \citet{bar98} and \citet{cha00b}.
We also include mass functions for the Pleiades \citep{mor04} and the
field \citep[][see Section~\ref{sec:field}]{boc10}.
As implied by the distributions of spectral types, the IMF in Taurus
peaks at a higher mass than the IMF in Chamaeleon~I and other young clusters.
Meanwhile, the mass function in Chamaeleon~I is roughly similar to those in
the Pleiades and the field. It is not possible to attribute small differences
between Chamaeleon~I, the Pleiades, and the field to true variations
in the IMF since each population is subject to different systematic errors.
Previous explanations for the unusually high peak mass in Taurus
have generally involved a higher average Jeans mass compared to other regions
\citep{bri02,luh03ic,luh04tau,goo04,lada08}. A recent numerical simulation
of the formation of both distributed and clustered stars has also reproduced
the variations in peak mass between Taurus and other clusters \citep{bon11a}.
However, the width of the IMF is predicted to be more narrow in a distributed
population than in clusters, whereas the IMF in Taurus is as broad as IMFs
in other regions. This suggests that the breadth of the IMF is not caused
by dynamical interactions.
As discussed in Sections~\ref{sec:open}, \ref{sec:sfr}, and \ref{sec:field},
surveys in the field and in young clusters indicate that the IMF
extends down to at least 0.01~$M_\odot$. A small number of objects 
have been identified that may have even lower masses, but they have
uncertain masses and sometimes are poorly characterized in terms of spectral
type and age. Thus, current data are insufficient for a stringent test
of the predictions of the minimum mass of the IMF.

To search for variations in the substellar mass function, the ratio of
the number of brown dwarfs to the number of stars is often compared among
young clusters \citep{bri02,luh03ic}. However, this ratio is probably
not a reliable diagnostic of such variations \citep{luh07cha}.
Because of the close proximity of the IMF's peak to the hydrogen burning
limit, the brown dwarf ratio is sensitive to true variations in the peak mass
(which would not provide any specific insight into brown dwarfs)
and false variations that stem from different choices
of evolutionary models, mass indicators (photometry vs. spectral type),
spectral classification schemes, and temperature scales.
This ratio also can depend on field size within a cluster because of
mass segregation \citep{mue03}.
Instead, it is better to characterize the low-mass IMF by its slope from
the peak down to the completeness limit when comparing results from different
studies since it is less susceptible to these systematic differences in
mass estimates.
As noted above, no significant variations in these slopes are apparent in
published IMFs for young clusters
\citep[$N_{BD}/N_{star}\sim0.2$,][references therein]{sle04,and06,luh07cha}.
Since the mass function of brown dwarfs is roughly the same regardless of
the presence of O stars, we can conclude that halting of accretion by
photoionizing radiation is not the dominant process that results in the
formation of brown dwarfs.
Similarly, Taurus has a very low stellar density, and yet it 
has produced a comparable abundance of brown dwarfs as much denser clusters
like the Trapezium, indicating that dynamical interactions are not essential
for the formation of brown dwarfs.
If brown dwarfs form predominantly in disks around solar-type stars,
then the numbers of brown dwarfs and solar-type stars should scale together,
but this is not seen in the comparison of Taurus to other clusters.
Indeed, given the paucity of solar-mass ($\sim$K7) stars in most clusters
(e.g., see IC~348 in Figure~\ref{fig:histo}), it seems unlikely that their
disks have hosted the formation of a large fraction of low-mass stars and brown
dwarfs. 

Finally, we note that the mass functions of pre-stellar cores in molecular
clouds resemble the stellar IMF in terms of the Salpeter slope at high
masses and the flattening at low masses, which has been taken as evidence
that gravitational or turbulent fragmentation determines the masses
of stars, and perhaps brown dwarfs \citep[][references therein]{rat09,and10}.
However, whether a physical relationship truly exists between these mass
functions remains a subject of debate \citep[e.g.,][]{smi09,cha10,mic11}.

\subsection{Multiplicity}

The properties of multiple systems are influenced by both the process of
star formation and dynamical interactions within and among stellar systems.
Thus, measurements of multiplicity may shed light on 
the formation and early evolution of low-mass stars and brown dwarfs.
Since this topic has been reviewed by \citet{bur07ppv},
we will focus on the new results since that review and the
implications of the observed multiplicity properties for the proposed
formation mechanisms.

\begin{figure}
\centerline{\psfig{figure=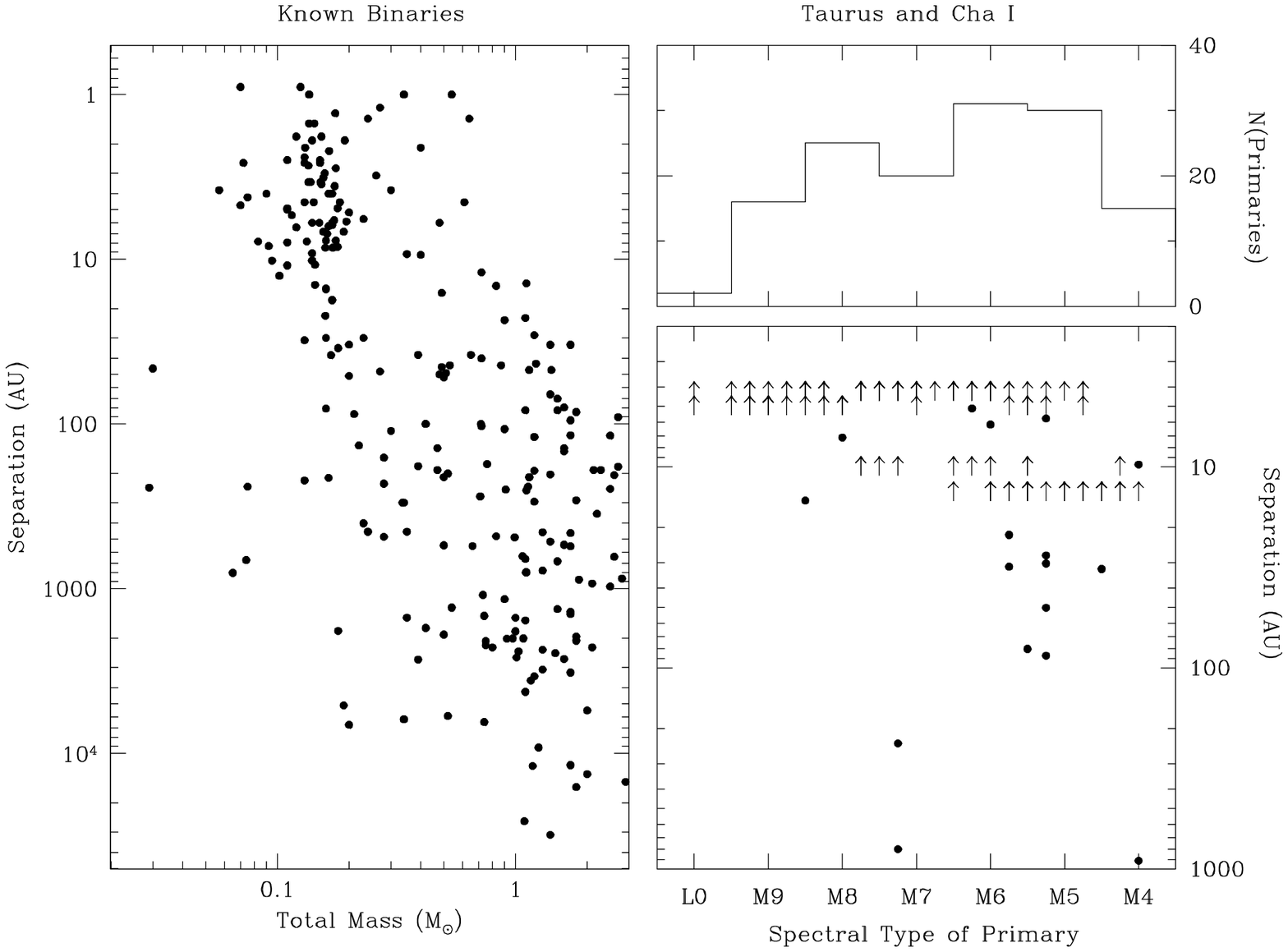,height=40pc}}
\caption{
{\it Left}: Projected separation versus total system mass for a compilation
of known binaries \citep{fah11a}.
{\it Right}: Projected separation of resolved binaries ({\it points}) and
the detection limits for unresolved sources ({\it arrows}) versus 
primary spectral type in Taurus and Chamaeleon~I
\citep[][K. Todorov, in preparation]{neu02,kra06,kon07,laf08}.
The number of primaries as a function of spectral type is shown with
the histogram.
}
\label{fig:bin}
\end{figure}

Multiplicity at low masses has been studied in both the solar
neighborhood and in young clusters. Each of the two populations has advantages.
Binary properties can be measured down to smaller separations in the field
while the youngest clusters allow us to characterize the
primordial multiplicity properties and their dependence on star-forming
conditions. For field stars and brown dwarfs, the binary fractions and
average separations decrease and the average mass ratios increase
($q\equiv M_2/M_1$) at lower primary masses
\citep[e.g,][Figure~\ref{fig:bin}]{dm91,fis92,bou03,bur03,clo03,bas06,ber10}.
The change in these properties with mass appears to be continuous
rather than abrupt \citep[e.g.,][]{kra11c},
resembling the behavior of the IMF, accretion rates,
and disk fractions (Sections~\ref{sec:imf}, \ref{sec:disks}).
Based on analysis by \citet{bur03}, it is unlikely that wide low-mass
binaries have been disrupted while in the field, and instead the
paucity of these systems is probably primordial or caused by dynamical
interactions at very young ages ($\tau\lesssim10$~Myr).
We can attempt to distinguish between these two possibilities by examining
the multiplicity of low-mass members of nearby star-forming regions
that have low stellar densities so that dynamical effects are minimized.
The best available options are Taurus and Chamaeleon~I. 
High-resolution imaging of the members of these regions has found
that the frequency of wide binaries and their average separation decrease
steadily from solar-mass stars to brown dwarfs \citep[][K. Todorov,
in preparation]{neu02,kra06,kra07a,kra11c,kon07,laf08,kra09a,kra11b}.
The trend at low masses is illustrated in Figure~\ref{fig:bin}, where
we plot the separations for binaries and the limits for unresolved sources
versus spectral type from M4 to L0.  The observed binary fractions
from these data are 13/62 ($21^{+6}_{-4}$\%) for M4--M6 and 5/74 
($7^{+4}_{-2}$\%) for $>$M6, which apply to separations of $>5$--10~AU.
Although the average separation decreases with later types, a few very wide
systems ($>100$~AU) are present at $>$M6 \citep{luh04bin,luh09fu}.
Therefore, we conclude that part of the dependence of multiplicity on primary
mass observed in the field is primordial, but dynamical interactions in denser
clusters are probably also responsible for the small frequency of very wide
binaries in the field. The role of environment can be examined by considering
additional star-forming regions. For instance, based on data similar to
those obtained in Taurus and Chamaeleon~I, members of Upper Sco later than 
M6 have an observed binary fraction of 0/22 \citep[$<8$\%,][]{kra05,bil11}.
This measurement is lower than the value in Taurus, 
but the difference is not statistically significant.

It would be desirable to closely compare multiplicity measurements between
the field and star-forming regions to search for signs of dynamical evolution
at ages of $\gtrsim10$~Myr.
For such a comparison to be meaningful, the selected star-forming region
should be representative of the birth places of most field stars.
It is likely that the field has been populated predominantly by
moderately rich clusters ($N>100$) within giant molecular clouds that
become OB associations (e.g, Orion, Upper Sco) rather than small, loose
aggregates \citep[e.g., Taurus,][]{lad03}.
However, the number statistics for the low-mass binary measurements in Upper
Sco are too low for a detailed comparison to the field, and measurements in
dense clusters like the ONC are hampered by chance alignments of unrelated
stars.

A few sets of formation models have made specific predictions for the
binary properties of brown dwarfs. We can compare those predictions to
observations, which can be summarized as follows based on the field
and cluster data: brown dwarfs have a binary fraction of 
20--30\% \citep{max05,bas06}, most of the binaries are tight ($<$20~AU) and
a few are very wide $>100$~AU, and the mass ratios tend to approach unity.
Early models produced too few binaries and no wide systems \citep{rei01,bat02}
but newer calculations are roughly consistent with the data
\citep{bat05a,bat09a,bat11,sta09}. The trends in mass ratio
and mean separation as a function of stellar mass are also 
qualitatively reproduced by \citet{bat09a,bat11}. However, it remains unclear
whether any of these models can make wide binary brown dwarfs in
the low-density conditions in which they have been primarily found.
The binary FU~Tau \citep[800~AU,][]{luh09fu} and the
quadruple containing 2MASS~04414489+2301513 \citep[1700~AU,][]{tod10}
would seem difficult to explain with dynamical or disk fragmentation
models given that they are both fragile and isolated.
Finally, the distribution of eccentricities predicted by \citet{bat09a}
is consistent with data for low-mass binaries in the field
while the binaries from \citet{sta09} are weighted too heavily toward
high eccentricities \citep{kon10,dup11}. However, the initial orbital
properties from these models could be altered by dynamical interactions,
making it difficult to conclusively test formation scenarios with data
from old binaries in the field.

\subsection{Circumstellar Disks}
\label{sec:disks}

Newborn stars experience much of their growth via accretion from
circumstellar disks, and it has been proposed that brown dwarfs arise
when this accretion is prematurely halted. As a result, observations of
accretion and disks around young low-mass stars and brown dwarfs may
prove useful in understanding their formation. 
As with multiplicity, this topic was reviewed in {\it Protostars and Planets V}
\citep{luh07ppv}, so we will provide a summary of the observations that
includes the latest results, and describe the implications for the
formation theories.

In the model of magnetospheric accretion, matter from a circumstellar
disk falls onto a young star along its magnetic field lines.
The impact of the material onto the stellar surface produces an accretion
shock, which radiates at UV and optical wavelengths.
The columns of material accreting onto a young star also emit hydrogen
emission lines that are significantly stronger and broader than those from
stellar chromospheres, offering an additional tracer of accretion.
Because H$\alpha$ is more easily observed in brown dwarfs than UV emission,
line profiles of H$\alpha$ provided the first evidence of accretion
for young low-mass objects \citep{muz00,jay02}.
Additional accretion diagnostics have included Ca~II \citep{moh05},
near-IR hydrogen transitions \citep{nat04}, and optical continuum veiling
\citep{whi03}. Although they are more difficult to collect, measurements
of UV excess emission have produced the most direct estimates of accretion
rates \citep{her08}. The resulting rates are correlated with those derived
from H$\alpha$, but are systematically larger by a factor of several.
The accretion rates from these various diagnostics are correlated with
stellar mass as $\dot{M}\propto M^2$ for the full range of masses
across which they have been measured 
\citep[$M\sim0.02$--2~$M_\odot$,][]{muz05,moh05,nat06,her08},
although it has been suggested that this apparent correlation
may be largely due to selection effects \citep{cla06}.  If a dependence on
stellar mass is present, its origin is unclear \citep{har06,vor09}. 
Accretion rates in low-mass objects (and young stars in general) exhibit
significant variability, which explains some of the scatter
in the correlation between accretion rate and stellar mass \citep{sch06b}.
As with stars, outflows should be a natural byproduct of accretion onto
brown dwarfs. Evidence of outflows from brown dwarfs has been detected in
the form of emission in optical forbidden lines and blue shifted absorption
in permitted lines \citep{fer01,whi04b,moh05,muz05}. A few of these outflows
have been spatially resolved through spectro-astrometry of the forbidden
lines \citep{whe05,whe09b} and submillimeter interferometry of CO
emission \citep{pha08,pha11}. Outflows also have been observed in
low-mass protostars, as discussed in Section~\ref{sec:proto}.

\begin{figure}
\centerline{\psfig{figure=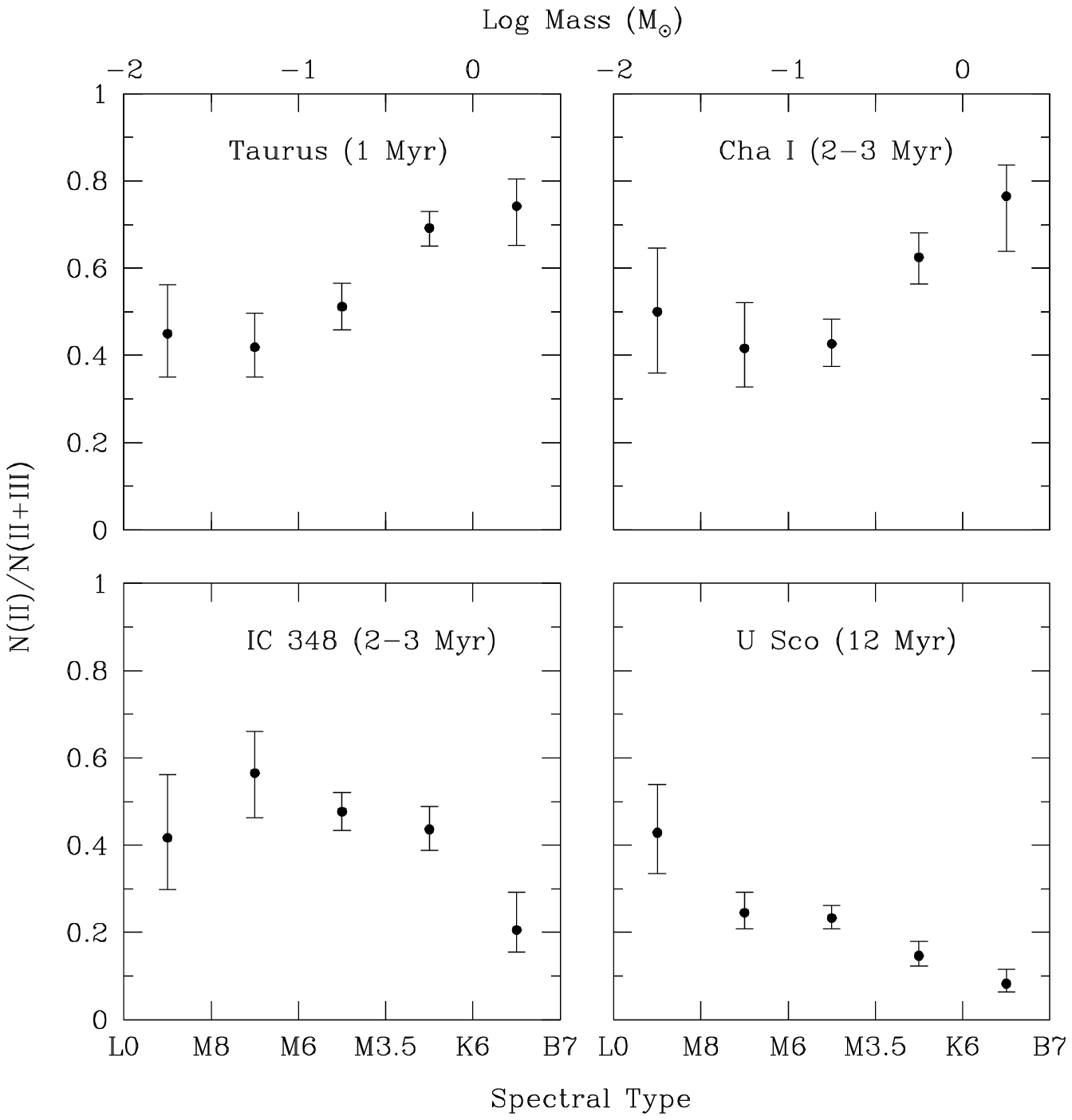,height=40pc}}
\caption{
Fraction of sources with primordial circumstellar disks (class~II)
as a function of spectral type in Taurus, Chamaeleon~I, IC~348, and Upper Sco
\citep[][K. Luhman, in preparation]{lada06,mue07,luh05frac,luh08cha1,luh10tau1}.
Stars with debris disks are designated as class~III.
The boundaries of the spectral type bins have been
chosen to correspond approximately to logarithmic intervals of mass.
}
\label{fig:disk}
\end{figure}

Circumstellar disks around young stars can be studied via their thermal
and line emission at IR and millimeter wavelengths where
they dominate the stellar photosphere.
Some of the earliest detections of disks around young objects near and
below the hydrogen burning limit were achieved through photometry
at 2--3~\micron\ \citep{luh99ic,mue01,jay03b,liu03}.
However, because of their low luminosities, most low-mass stars and brown dwarfs
do not heat their disks sufficiently for significant excess emission to
appear at those wavelengths \citep{luh10tau1}. As a result, photometry at
longer, mid-IR wavelengths is necessary for reliable detections of these
relatively cool disks.  The {\it Infrared Astronomical Satellite} (IRAS)
measured mid-IR photometry for a few low-mass stars
in sparse regions like Taurus \citep{kh95} and similar data were
collected for some of the brightest young brown dwarfs by {\it ISO}
\citep{com98,nat02} and large ground-based telescopes \citep{apa04,ster04}.
The arrival of the {\it Spitzer Space Telescope} resulted in a breakthrough
in surveys for disks around brown dwarfs (as well as stars in general).
The unprecedented mid-IR sensitivity of {\it Spitzer} enabled
detections of disks around the faintest and coolest known members
of the nearest star-forming regions \citep{luh05cha2}.
In addition, it could perform these measurements for nearly all of the
members of these regions because of its ability to efficiently map large
areas of sky \citep[][references therein]{luh10tau1}.
The data from {\it Spitzer} have been used to measure the fraction
of stars and brown dwarfs that harbor inner disks as a function of
spectral type, which acts as a proxy for stellar mass. Figure~\ref{fig:disk}
shows some of the best available disk fractions in terms of number
statistics and completeness at low masses, which apply to Taurus,
Chamaeleon~I, and IC~348 \citep{lada06,mue07,luh05frac,luh08cha1,luh10tau1}.
We also include new disk fractions for Upper Sco based on WISE photometry
\citep[K. Luhman, in preparation, see also][]{car06}.
In these data, disks remain common in star-forming regions down
to the least massive known members ($\sim0.01$~$M_\odot$). 
In fact, the evolution of these disk fractions with age from the youngest
to oldest region indicates that the lifetimes of disks around low-mass 
stars and brown dwarfs are longer than those of more massive stars.
The first hints of the potential longevity of disks at low stellar masses
were provided by the detections of disks around two of the brown dwarfs in TWA
\citep{moh03b,ster04,mor08,ria08}.

A number of studies have sought to characterize the physical properties of
the circumstellar disks identified in the {\it Spitzer} surveys.
The radii of most disks around young stars cannot be directly measured
with current technology. However, if a disk is viewed nearly edge-on, it
can occult the star and be detected in scattered light. Using a combination
of {\it Spitzer} spectroscopy and high-resolution imaging from {\it HST},
\citet{luh07edgeon} detected an edge-on disk around a young brown dwarf in
Taurus and were able to estimate a disk radius of 20--40~AU. 
SEDs of young stars do not normally provide good constraints on disk radii
because of degeneracies between radius and other parameters in disk models,
but the circumprimary disk of the brown dwarf 2MASS~J04414489+2301513
appears to be so small ($R=0.2$--0.3~AU) that the degeneracies are minimal
\citep{ada11}. The small size of this disk may be due to truncation by
its secondary \citep[$M\sim5$--10~$M_{\rm Jup}$, $a=15$~AU,][]{tod10}.
No other measurements of disk radii are available for young brown dwarfs.
Masses have been measured for small samples of young low-mass objects
through far-IR and millimeter observations \citep{kle03,sch06a,harv11,moh11},
which have produced values of $\lesssim1$~$M_{\rm Jup}$ assuming a standard
gas-to-dust ratio of 100.
Based on modeling of SEDs, the geometries of brown dwarf disks range from
flared to relatively flat \citep{nat01,moh04c,apa04},
which are believed to correspond to varying degrees of dust grain growth
and settling. {\it Spitzer} spectroscopy of the 10~\micron\ silicate emission
feature has provided direct evidence of dust processing and the production of
crystalline grains \citep{fur05,apa05}.
The silicate feature is weaker at later spectral types
\citep{kes07,sch07a,mor08,fur11}, indicating that grain growth may occur faster
in disks around brown dwarfs than in disks around stars. 
The disks of low-mass stars and brown dwarfs may also differ from those of
solar-type stars in terms of their abundances of organic molecules
\citep{pas09}. The evolution of disks may depend on stellar mass as well;
disks with inner holes, known as transitional disks, have been uncovered
around low-mass objects \citep{muz06,fur11}, but it appears that they
are less common at lower stellar masses \citep{muz10}.

In all of the proposed formation mechanisms from Section~\ref{sec:theory},
brown dwarfs harbor accretion disks when they are born.
Some of the models suggest that brown dwarfs have low masses because dynamical
interactions ended accretion early, but few concrete predictions
have been made regarding disk properties. As a result, all of the models
are technically consistent with all of the observations of brown dwarf disks,
such as the continuity of accretion rates and disk fractions from solar-type
stars to brown dwarfs.
The ejection models do predict that brown dwarf disks will have radii of
$\lesssim10$~AU \citep{bat03}, but they allow for a few somewhat larger disks
\citep{bat09a}, and thus are consistent with the 20--40~AU disk in the
edge-on system mentioned above. The measurements of disks around
low-mass stars and brown dwarfs described in this section provide
valuable constraints on the stellar mass dependence of the physics of
accretion, disk evolution, and planet formation, but unfortunately they
have not offered direct tests of the possible formation mechanisms for
brown dwarfs.

\subsection{Protostars}
\label{sec:proto}

All of the proposed formation mechanisms for brown dwarfs occur during
either cloud fragmentation or the protostellar phase. 
Thus, observations of the youngest progenitors of low-mass stars
and brown dwarfs should provide direct constraints on the formation models.
For instance, brown dwarfs that form by disk fragmentation or
ejection should never appear as isolated protostars.

A defining characteristic of a protostar is the presence of both an
accretion disk and an infalling envelope. Protostars are often
designated as class 0 or class I based on their SEDs,
where class 0 is redder and presumably less evolved than class I
\citep{lada87,and93}. The all-sky mid-IR data from IRAS enabled
the first comprehensive surveys for protostars, which uncovered a few class~I
candidates with spectral types as late as $\sim$M5 ($M\sim0.15$~$M_\odot$),
such as IRAS~04158+2805, IRAS~04248+2612, and IRAS~04489+3042 in Taurus
\citep{kh95}. The class~I nature of these objects has been confirmed by the
detections of envelopes \citep{whi04b,and08,fur08}.
High-resolution images from {\it HST} of the protostar IRAS~04325+2402
have revealed a faint nebulous companion that could be a class~I brown dwarf
\citep{har99,sch08b}, although confirmation via mid-IR observations is
not currently possible because of its small separation from the primary.

To identify free-floating protostellar brown dwarfs,
sensitive wide-field mid-IR images are necessary, and the first data of this
kind became available via the {\it Spitzer Space Telescope}, as mentioned
previously in this review.
Surveys for low-mass protostars have been conducted by either obtaining
{\it Spitzer} images of samples of molecular cloud cores, often selected
based on the absence of previously detected stars (``starless" cores), or
searching for
candidate protostars anywhere in the {\it Spitzer} images of a star-forming
region regardless of the availability of previous detections of cores
\citep{dun08}. In both cases, the candidates must be confirmed as protostars
associated with the target molecular cloud. For class~0 sources, a commonly
employed signature of cloud membership is the presence of an outflow
\citep{bou05}. Since they have less obscuration, class~I candidates are
often bright enough at near-IR wavelengths for spectroscopy to verify that
they are young brown dwarfs rather than galaxies \citep{luh10tau2}.
It is also necessary to detect the envelopes of class~I objects since
the SEDs of class~II sources (star+disk) can resemble class~I SEDs
if they are highly reddened, as in the case of the Taurus brown dwarf
2MASS~J04194657+2712552 \citep{luh09tau,fur11}.
To date, {\it Spitzer} surveys have identified more than a dozen candidate
class~0 sources with luminosities that are indicative of young brown dwarfs
\citep[$L<0.1$~$L_\odot$,][]{dun08}, several of which have been 
confirmed as protostellar \citep{you04,bou06,dun06,dun10,lee09,kau11}.
Some of these objects may eventually accrete enough to become stars
\citep{dun10} while others seem destined to remain substellar given their small
accretion reservoirs \citep{lee09}.
The existence of isolated protostellar brown dwarfs would indicate that
the dynamical interactions, fragmentation in disks, and photoionizing
radiation from massive stars are not essential for the formation of
free-floating brown dwarfs.

\subsection{Kinematics and Positions at Birth}
\label{sec:spatial}

Since dynamical interactions play a central role in some of the proposed
formation mechanisms for low-mass stars and brown dwarfs, the observed
kinematic and spatial distributions of these objects can provide tests of the
models. For instance, some of the earlier ejection models predicted that brown
dwarfs are born with higher velocities than stars \citep{rei01,kro03b}.

Radial velocities have been measured for only small samples
of late-type members of star-forming regions, but the
available data indicate that young stars and brown dwarfs have similar
velocity dispersions \citep{joe01,whi03,joe06,kur06}.
Brown dwarfs also share the same spatial distributions as stars in Taurus and
Chamaeleon~I, whereas they should be more widely distributed if they are
born with higher velocities \citep{gui06,sle06,luh06tau1,luh07cha,park11}.
These results are consistent with most of the latest theories for the
formation of brown dwarfs.

Both the ejection models and the disk fragmentation models require
the presence of stars to produce brown dwarfs since stars facilitate
the dynamical interactions that lead to ejection and act as the hosts of
the fragmenting disks.
As a result, it should not be possible for brown dwarfs to form in isolation if
either of these mechanisms is required for their formation.
Taurus is the best site for identifying brown dwarfs that are likely
to have been born alone, and thus testing this prediction, since it is one of
the richest nearby star-forming regions ($\sim400$ members) as well as one
of the most sparsely distributed ($n\sim 1$-10~pc$^{-3}$).
The components of the wide binary FU~Tau are the clearest examples of brown
dwarfs in Taurus that have been born in relative isolation \citep{luh09fu}.
They are projected against the center of the Barnard 215 dark cloud,
which is their likely birth place.
Only one other known member of Taurus, FT~Tau, is found within $0.5\arcdeg$
of this system. The fact that FU~Tau~A and B were born in the absence of
stars indicates that dynamical processes and disk fragmentation played no
role in their formation. The probable protostellar brown dwarfs
described in Section~\ref{sec:proto} also have few neighboring stars
and may be younger analogs of FU~Tau.

\section{EARLY STELLAR EVOLUTION AT LOW MASSES}

\subsection{Theory}

The early evolution of low-mass stars and brown dwarfs ($\tau\lesssim100$~Myr)
can be characterized largely in terms of the change in bolometric
luminosity, effective temperature, radius, and angular momentum over time
for a given stellar mass.
Theoretical models of stellar interiors and atmospheres have made
predictions for the evolution of the first three parameters while other
structures (e.g., winds, disks) may lead to loss of angular momentum.
The basic ingredients of modern evolutionary and atmospheric models 
for low-mass stars and brown dwarfs have been reviewed
by \citet{all97} and \citet{cha00c}. Over the last decade,
the possible effects of additional processes have been explored
and some of the major results can be summarized as follows:

\begin{enumerate}

\item
Magnitudes and colors produced by evolutionary models are sensitive to
uncertainties in molecular line lists and opacities, but luminosities
and temperatures are not affected significantly \citep{bar02}.

\item
Evolutionary tracks on the H-R diagram for a given mass are sensitive
to the initial radius for ages of $\lesssim1$~Myr \citep{bar02}.

\item
The predicted temperature for a given mass and age and the change in
temperature over time (i.e., verticality of mass tracks) depend on the
choice of mixing length, particularly at younger ages \citep{bar02}

\item
The radius and luminosity may be influenced by previous episodes of intense
accretion for ages up to $\sim10$~Myr. The direction of this change is
determined by whether thermal energy from the accreting matter is added
to the star \citep{har97,bar09,hos11}.

\item
The presence of strong magnetic fields and heavily spotted surfaces
may result in a larger radius and a cooler temperature \citep{cha07,mac09}.

\item
Unstable modes with timescales of 1--5~hours may occur during deuterium
burning \citep{pal05a}.

\end{enumerate}

To test the accuracy of the standard models for the early evolution of low-mass 
objects in light of the above possible complications,
we describe in this section observational constraints on the evolution
of angular momentum, temperature, luminosity, and radius from $\sim$1--100~Myr
provided by measurements of rotation, dynamical masses, and empirical
isochrones on the H-R diagram.

\subsection{Rotation}

Much of the early work on the angular momentum evolution of young stars
focused on rotational velocities measured from high-resolution spectra
of absorption line profiles \citep{har86,bou86,stau86}.
Because of the development of wide-field optical cameras in the 1990's,
it became more efficient to trace angular momentum via rotation periods (in
compact clusters), which can be measured from the photometric modulations
caused by the rotation of spotted stellar surfaces.
Rotation periods are now available for thousands of cluster members, including
an increasing number of low-mass stars and brown dwarfs \citep{her07}. 

For stars at ages of a few Myr, the largest studies of rotation periods have
been conducted in the ONC and NGC~2264
\citep[$\tau\sim1$ and 2~Myr,][]{stas99,her01,her02,lam04,lam05}.
In any given mass range, stars in these clusters exhibit a broad 
distribution of periods, and hence angular momenta, covering a
factor of 30 or more. The origin of a breadth of this kind at such young ages
is unknown since rotation measurements at earlier, protostellar stages
are difficult to obtain \citep{cov05}.
For masses of 0.4--1.5~$M_\odot$, the periods range from 0.6--20 days
and have a bimodal distribution. 
Stars at lower masses rotate more rapidly on average and show a single
peak in their period distribution.
Although rotation periods are shorter at lower masses, the specific
angular momentum remains roughly constant.
These trends appear to continue into substellar masses, where young brown
dwarfs rotate with periods of $\lesssim1$ day \citep{sch05a,rod09,cod10}.
\citet{sch05a} suggested that some young low-mass objects rotate near 100\%
of the break-up period based on a small sample in $\epsilon$~Ori, but
\citet{cod10} found that the maximum rotation period is 40\% of break-up
in a larger sample in $\sigma$~Ori.
The fraction of low-mass stars and brown dwarfs that show periodic variability
is lower than the fraction for solar-type stars, probably because of weaker
magnetic fields and fewer spots \citep{cod10}.
In addition to measuring rotation periods, the photometric monitoring
of \citet{cod10,cod11} found no evidence that young brown dwarfs
undergo pulsations from the instabilities examined by \citet{pal05a}.

To explain the presence of slowly rotating stars at ages of only a few Myr,
it has been proposed that a young star can lose angular momentum
by transferring it to a circumstellar disk via the stellar magnetic field
\citep[``disk locking",][references therein]{her07}.
If this mechanism is valid, then slow rotators should show a higher
frequency of disks. This prediction was tested initially with ground-based
near-IR measurements of disk excesses, but mid-IR data from
the {\it Spitzer Space Telescope} have enabled more conclusive tests,
particularly for low-mass stars and brown dwarfs, whose disks
often lack noticeable near-IR excesses.
Among solar-type stars, slow rotators do show 
a higher frequency of disks, supporting the disk locking theory in this
mass range \citep{reb06,cie07}.
For low-mass stars, some studies have found evidence for a
connection between rotation and disks \citep{reb06}
while others have not \citep{cod10,leb11}.
Thus, disk locking may be less efficient at lower stellar masses.

Rotation measurements have been compared among clusters across a wide
range of ages to characterize the degree of angular momentum loss over time.
Solar-type stars rotate progressively faster on average from the ONC to
NGC~2264 to the zero age main sequence in the Pleiades, IC~2602, and
$\alpha$~Per \citep{her05}. The specific angular momentum changes little
over time for the faster rotators while stars that rotate slowly 
at ages of a few Myr seem to experience additional breaking.
Meanwhile, the evolution of low-mass stars is more closely described
by conservation of angular momentum with only exponential wind breaking
\citep{sil00,ter00,sch04b}.
Few rotational data are available for brown dwarfs near ages of $\sim$100~Myr
for comparison to younger counterparts.
Measurements for older brown dwarfs in the field
indicate that they do eventually spin down, but on much longer
timescales than stars \citep{moh03a,zap06,rei08,bla10,irw11,sch11}.
This trend of decreasing angular momentum losses at lower masses has
been attributed to changes in the mechanisms of magnetic field generation
from solar-type stars to fully convective low-mass stars and brown dwarfs and
the reduced coupling of atmospheres and magnetic fields at lower temperatures.

\subsection{Dynamical Masses}
\label{sec:dyn}

The luminosity, temperature, and radius predicted by evolutionary models
for a given mass and age can be tested by comparison to the observed
properties of stars and brown dwarfs whose masses are known independently
from dynamical measurements. Binary systems are the most common source
of dynamical masses, but it also has been possible to estimate the masses of
a few young stars via the rotational velocities of their circumstellar
disks \citep{sim00}.
Dynamical masses have been measured for a few dozen pre-main-sequence stars, 
and compilations of these data have been used to characterize the sizes of
systematic errors in various sets of models and the possible sources of
those errors \citep{bar02,luh03ic,hil04,mat07ppv}.

We now examine the model constraints provided by the small number of
young low-mass stars and brown dwarfs for which dynamical data are available.
These objects consist of two eclipsing binaries in the
ONC, JW~380 \citep[0.15 and 0.26 $M_\odot$,][]{irw07} and
2MASS~J05352184-0546085 \citep[0.0366 and 0.0572 $M_\odot$,][]{stas06,gom09}, 
the spectroscopic binary PPl~15 \citep[$q=0.85$,][]{bas99}, and the resolved
companion AB~Dor~C \citep[0.09~$M_\odot$,][]{clo05,gui06b}.
The measured radii of the components of JW~380 are consistent with the
values predicted by most models for an age of a few Myr \citep{irw07}.
We have measured a spectral type of M4.5 from an optical spectrum of the
system. When combined with the temperature scale
from \citet{luh03ic} and the 1~Myr isochrone of \citet{bar98}, this
classification implies a mass of $\sim0.2$~$M_\odot$, which falls
between the dynamical masses of the components as it should.
Thus, the data for JW~380 suggest that the IMFs for star-forming clusters
that are based primarily on spectral types, like those in Figure~\ref{fig:imf},
do not have large systematic errors near 0.2~$M_\odot$. JW~380 also
confirms that the peak in the spectral type distributions for star-forming
regions corresponds to a mass of $\sim0.15$~$M_\odot$ (Figure~\ref{fig:histo}).

Like JW~380, the radii of the components of 2MASS~J05352184-0546085
(hereafter 2M~0535-0546) are in rough agreement with the predictions of
evolutionary models \citep{stas06,stas07}. 
A spectral type of M6.5 has been estimated for the primary \citep{stas06}
while we classify a spectrum of the combined system as M6.75. 
Thus, this system has confirmed that newborn brown dwarfs are as large
and warm as old low-mass stars. However, the eclipse data have shown that
the primary is warmer than the secondary, which is contrary to the expectations
of standard evolutionary models.
It has been suggested that the primary may be unusually cool (and large)
for its mass because of a strong surface magnetic field and/or a large
spot coverage \citep{stas07,cha07,mac09}.
The primary rotates faster than the secondary and exhibits stronger
chromospheric H$\alpha$ emission \citep{rei07b}, which suggests that
it is indeed magnetically active.
Low-mass stars in tight eclipsing binaries in the field also tend to be
cooler and larger than expected based on theoretical predictions
\citep[][references therein]{mor09}. A magnetic origin for this anomaly
is supported by recent studies indicating that the radii of stars in 
wider binaries are less inflated relative to the models \citep{kra11a,cou11}
and that single rapid rotators and active stars may be cooler
and larger than normal \citep{mor08b,jac09}.
If this explanation for the temperature reversal of 2M~0535-0546 is correct,
then the spectral types of young brown dwarfs may be altered by the
presence of activity, leading to errors in the mass estimates of individual
objects \citep{moh09}. We discuss the implications of this source
of error on measurements of the IMF in Section~\ref{sec:imp}.
Additional tests of the evolutionary models at young ages and low masses
should soon be available via new eclipsing binaries that have been
recently discovered in the ONC
\citep[][Morales-Calder\'{o}n, in preparation]{mor11}.

Since the age of AB~Dor is similar to that of the Pleiades
\citep[$\tau=75$--150~Myr,][]{luh05ab}, it is useful to discuss AB~Dor~C and
PPl~15 together. As with JW~380 and 2M~0535-0546, we can use AB~Dor~C and
PPl~15 to help calibrate the relation between mass and spectral type,
but now at an age of $\sim100$~Myr.
AB~Dor~C has been classified as M5.5--M6 \citep{luh06ab,clo07b} and
the components of PPl~15 likely have spectral types of M6 and M7 \citep{bas99}.
Since the dynamical measurements of PPl~15 have provided only a mass ratio
rather than individual masses, we instead make use of the upper mass limit
of 0.1~$M_\odot$ for each component derived from the presence of Li
\citep{bas96,bas99}. The mass constraints for AB~Dor~C and PPl~15
combined with their spectral types suggest that the hydrogen burning
limit occurs near M6 at ages of $\sim100$~Myr, which is consistent with
the temperature predicted by evolutionary models \citep[e.g.,][]{cha00b}
for a reasonable choice of temperature scale \citep{luh99ic}.
The near-IR absolute magnitudes of AB~Dor~C are also consistent with
the values predicted for its age \citep{boc08}.

Finally, we note that significant progress has been made over the
last few years in measurements of dynamical masses for low-mass stars
and brown dwarfs in the field \citep{kon10,dup10}.
These data are valuable for testing models of older brown dwarfs
($\tau\gtrsim0.5$~Gyr), but the results of those tests are not directly
relevant to the early evolution of low-mass objects since the likely sources
of error in the models differ between different regimes of temperature and age.

\subsection{Hertzsprung-Russell Diagram}
\label{sec:hr}

Empirical isochrones formed by the members of multiple systems
and young clusters have long been used to assess the validity of 
theoretical isochrones. This test has been applied to young low-mass
objects in star-forming regions down to spectral types of M9
\citep{whi99,luh04bin,kra09b}, revealing relatively good
agreement between the empirical isochrones and some sets of models
\citep{bar98}, particularly when flexibility in the temperature
scale is allowed \citep[][Section~\ref{sec:teff}]{luh99ic,luh03ic}.
In this section, we expand this work to later types and older ages
by considering the best available samples of objects later than M6 and
younger than $\lesssim100$~Myr. By doing so, we hope to
more fully illuminate the evolutionary paths taken by young, low-mass objects.

To display the early evolution of low-mass stars and brown dwarfs, we
plot objects on a diagram of $M_{K_s}$ versus spectral type. We use
an H-R diagram in terms of observed properties rather than temperature
and luminosity to preserve the data in their original form and to
avoid the uncertainties associated with estimates of temperature and
luminosity. Spectral type is the best observational proxy for temperature,
and it can be measured quite precisely with the proper data and methods.
Among the standard broad-band filters, we use $M_{K_s}$ as a substitute
for luminosity because, in this band, excess emission from disks around
late-type objects is negligible and extinction is low. $M_{K_s}$ should also be 
a better choice than bands at shorter wavelengths since the near-IR colors of
young objects become unusually red at the latest types (Section~\ref{sec:bc}).

\begin{figure}
\centerline{\psfig{figure=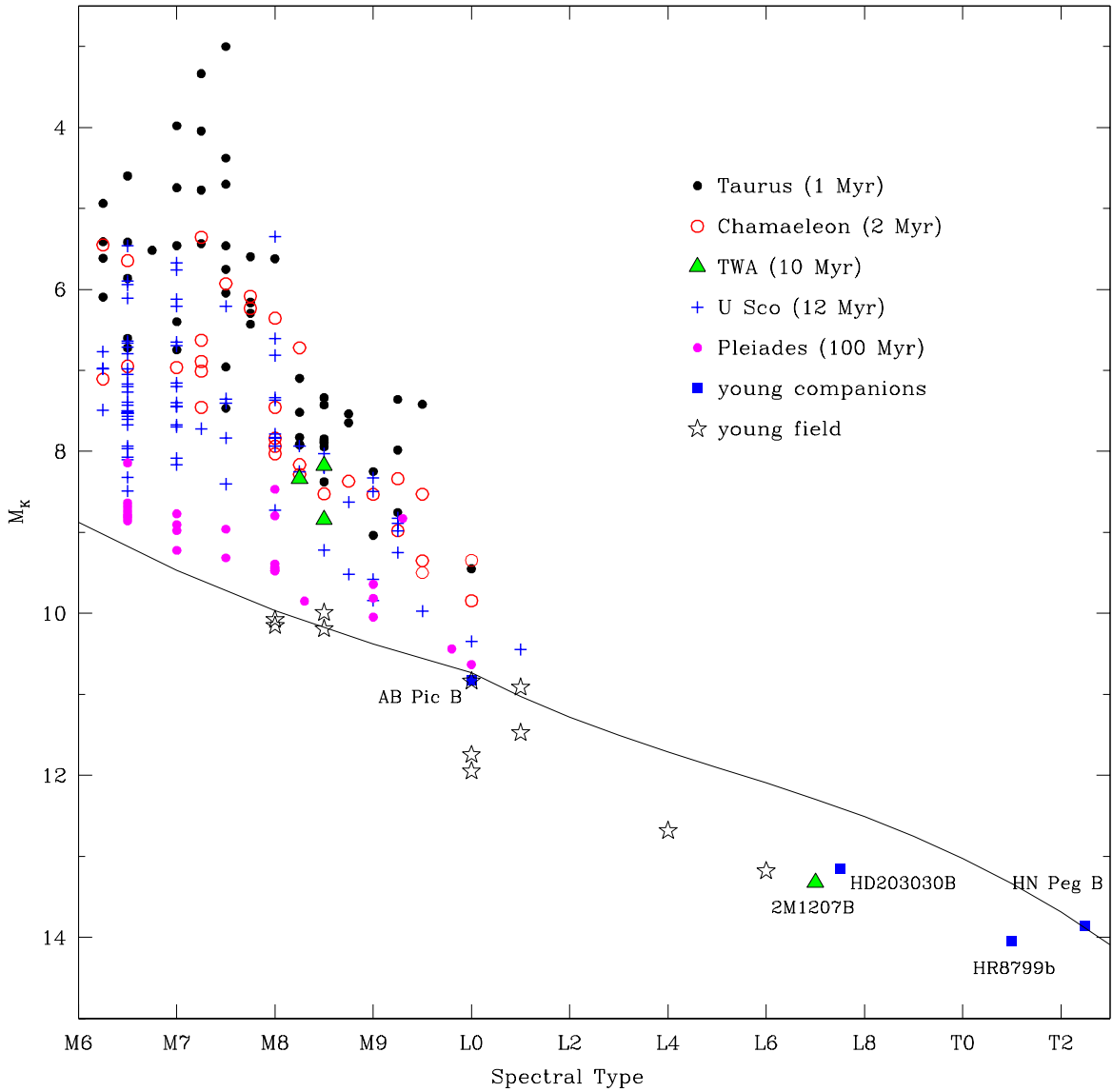,height=40pc}}
\caption{
$M_K$ versus spectral type for young late-type objects in nearby 
associations, clusters, multiple systems, and the field. A fit to data for
normal, older field dwarfs is shown \citep[{\it solid line},][]{fah11b}. 
}
\label{fig:hr}
\end{figure}

For our H-R diagram, we have selected young populations in which
the known members extend to spectral types of $\gtrsim$L0.
This sample consists of Taurus ($\tau\sim1$~Myr), Chamaeleon
($\tau\sim2$--3~Myr), TWA ($\tau\sim10$~Myr), Upper Sco ($\tau\sim12$~Myr),
and the Pleiades ($\tau\sim100$--125~Myr, see Sections~\ref{sec:open},
\ref{sec:sfr}).
We also include young field dwarfs and young companions for which
accurate spectral types and parallaxes are available. The former
are from \citet{fah11b} and the latter consist of the previously discussed
companions 2M~1207-3932~B, HR~8799~b, and HN~Peg~B, as well as AB~Pic~B
\citep{chau05b,bon10} and HD~203030~B \citep{met06}.
G~196-3~B is another companion that is cool, young, and well-studied
\citep{zap10}, but it is excluded since it lacks a parallax measurement.
Similarly, we consider only the members of TWA that have parallax data.
The H-R diagram for these populations is shown in Figure~\ref{fig:hr}.
We have attempted to ensure that the adopted spectral types are on
the same classification system, and have reclassified a few
objects with spectra from the literature and our unpublished data.
We have also omitted sources that appear to have uncertain spectral types
or membership. Most of the photometry is from 2MASS and UKIDSS, and
no correction has been made for known binaries that are unresolved in those
images. For comparison, we include in Figure~\ref{fig:hr} a fit to 
the sequence of M, L, and T dwarfs in the field \citep[$\tau>1$~Gyr,][]{fah11b}.

In analyzing the cluster sequences, we begin by examining their thicknesses
in $M_{K_s}$. The sequences for Taurus, Chamaeleon, and Upper Sco have widths of
1--2~mag in $M_{K_s}$ at a given spectral type while the Pleiades sequence
is much more narrow, and most of its bright outliers are known binaries.
Large vertical spreads are universally seen at higher mass as well
for star-forming regions, and have been primarily attributed to either
a wide distribution of ages or observational uncertainties
\citep{har01,pal02}.
The latter should be minimized in our H-R diagram since the photometry
is fairly accurate in most cases, extinction errors are small in the $K_s$
band, many of the spectral types have been measured in the same manner, 
disk emission is negligible, and there is little if any contamination
from non-members. Unresolved binaries can introduce a spread of 0.75~mag
and the large diameters of Taurus and Upper Sco correspond to a range
of $\sim$0.35~mag in the distance moduli of their members.
The sequences seem somewhat thicker than expected from these effects alone,
so it is possible that a significant spread of $M_{K_s}$ is truly present.
If so, it may reflect a spread in radii rather than ages.
For instance, the ONC appears to exhibit a wide distribution of radii that
is not attributable to ages \citep{jef07,jef11}, and instead may be caused by
differences in accretion histories \citep{bar09,lit11}, although this
interpretation has been challenged \citep{hos11}.
The observation that members of binaries in Taurus are more coeval on the
H-R diagram than random pairs of Taurus members suggests that at least
a small portion of the luminosity spread in this region is due to a range
of ages \citep{whi01,kra09b}.

The comparison of cluster sequences in Figure~\ref{fig:hr} provides
a test of some of the basic predictions for the early evolution of low-mass
objects. For instance, theoretical isochrones and mass tracks
converge into one narrow sequence below $\sim$1500~K since at
these temperatures all brown dwarfs are predicted to have similar radii,
even the younger ones \citep{bur97,bar98,cha00b}. Indeed, this feature is
evident in Figure~\ref{fig:hr}, where the cluster sequences steadily
converge from earlier to later spectral types, eventually approaching the
sequence for field dwarfs. Although the temperature scale for young brown
dwarfs is uncertain, particularly at later types (see below), it is useful to
attempt a quantitative comparison to the predicted luminosity evolution.
From Taurus to the Pleiades, the median sequences fade by
$\sim3.5$ and 1~mag in $M_{K_s}$ at M7 and L0, respectively,
which is consistent with the predicted changes in luminosity from
1--100~Myr for 2900 and 2200~K.

The behavior of the coolest objects in Figure~\ref{fig:hr} is quite peculiar.
With the exception of the members of star-forming regions, 
young sources later than M9 fall on or below the sequence for field dwarfs,
which is unexpected since young brown dwarfs should always be
brighter than older ones at a given temperature.
A number of studies have separately stumbled across the underluminous
nature of objects in Figure~\ref{fig:hr}.
Based on their unusually red colors (Section~\ref{sec:bc}) and their
temperatures estimated from model spectra, 2M~1207-3932~B and the companions
to HR~8799 appear below theoretical isochrones for their expected ages
\citep{moh07,bow10}.
HD~203030~B and HN~Peg~B have been reported to be cooler than field dwarfs
with the same spectral types \citep{met06,luh07tdwarfs}, which is equivalent
to saying that, if they have the same temperature as field dwarfs with the
same types, then they are too faint for their youth (i.e., underluminous).
The presence of this common anomaly has been recognized for subsets
of these objects \citep{met09b,bow10,cur11,ske11}.
\citet{met09b} and \citet{bow10} attributed it to a gravity dependence of
the temperature of the L/T transition, but Figure~\ref{fig:hr} shows
that earlier L dwarfs appear underluminous as well.
Meanwhile, \citet{cur11} suggested that the origin of this property
is related to the planetary nature of the companions.
Through their work on young field dwarfs, \citet{fah11b} realized
that these young companions and young L dwarfs are both unusually red
in near-IR bands and appear to be underluminous, concluding that
their youth is the source of these characteristics.
We now add to the objects that follow this pattern by noting that
the sequences for young clusters like Taurus and Chamaeleon become
underluminous relative to theoretical isochrones for spectral types
later than M9 \citep{luh08cha2,luh09tau}.

The underluminous positions of young objects in Figure~\ref{fig:hr}
and other varieties of H-R diagrams can be partially explained by
the fact that the near- to mid-IR colors of young objects later than M9 
are redder than those of their field counterparts, and thus do not
have the same bolometric corrections (Section~\ref{sec:bc}).
It is also likely that the temperature scale for young L and T dwarfs
is significantly cooler than that of field dwarfs.
The latter explanation is supported by the work of \citet{bar11a,bar11b},
who were able to successfully fit the SEDs of HR~8799~b and 2M~1207-3932~B
with models of low-gravity, cloudy atmospheres that experience non-equilibrium
chemistry. According to their models, a much lower temperature is needed
for methane absorption to appear in young objects than in field dwarfs.
As a result, \citet{bar11b} suggested that only extremely low-mass members
of star-forming regions should be cool enough to exhibit methane and appear
as T dwarfs.

Although the data in Figure~\ref{fig:hr} reveal a systematic pattern in
the photometric properties of young L dwarfs, we point out one anomalous
aspect. All of the young L dwarfs from the field have gravity-sensitive lines
that are indicative of ages of $\lesssim100$~Myr, so it is unclear
why two of the L0 dwarfs appear well below the Pleiades sequence. 
In fact, one of these objects, 2MASS~J00325584-4405058, has been classified
as L0$\delta$, which suggests an age of $\sim10$~Myr \citep{cru09}.
A similar, yet smaller, discrepancy is present for AB~Pic~B, which
is believed to have an age of 30~Myr but is slightly fainter than its
counterpart in the Pleiades. It would be interesting to directly compare
spectra of the various objects near L0 to verify that they have the
same spectral types, determine their relative surface gravities, and
assess whether the latter are consistent with their relative
absolute magnitudes.

\subsection{Accuracy of IMF Measurements in Young Clusters}
\label{sec:imp}

\begin{figure}
\centerline{\psfig{figure=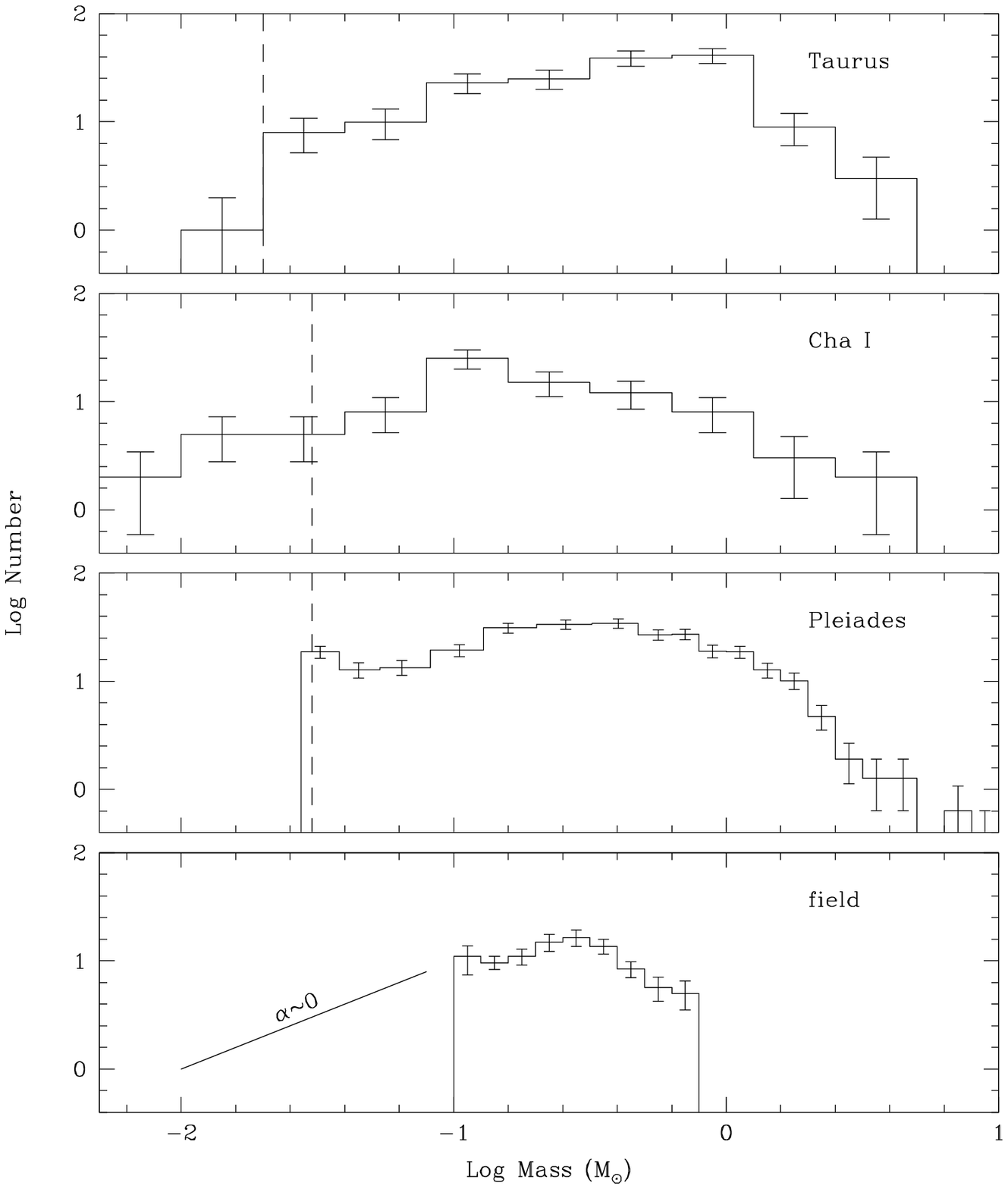,height=40pc}}
\caption{
IMFs for Taurus, Chamaeleon~I, the Pleiades, and the field
\citep{mor04,luh07cha,luh09tau,boc10}. Assuming a power-law form for
the mass function of brown dwarfs, surveys of the field indicate a slope
of $\alpha\sim0$ in linear units, or $\Gamma=-1$ in the logarithmic
units plotted in this diagram \citep{met08,burn10,rey10,kir11}.
The data for the Pleiades and the field have been scaled to fit within
the same limits used for Taurus and Chamaeleon~I.
The completeness limits of the cluster samples are indicated
({\it dashed lines}). 
}
\label{fig:imf}
\end{figure}

Given the possible errors in evolutionary models described in the
previous sections, it is useful to discuss the resulting implications
for IMF estimates in young clusters.
Since very few cluster members have been spectroscopically confirmed at types
later than M9, the errors in bolometric corrections and temperature scales
described earlier for young L dwarfs have little impact on previous IMF
measurements. For the IMFs in the youngest clusters (e.g, Figure~\ref{fig:imf}),
the masses are derived from positions in the H-R diagram, and hence
depend very little on luminosity. As a result, processes that primarily
affect luminosity, such as episodic accretion, should not introduce large
errors into the mass estimates. On the other hand, the spectral types of young
low-mass stars and brown dwarfs may be altered by magnetic activity
(Section~\ref{sec:dyn}).
To gauge the magnitude of this source of error, we can consider the width
of the peak in the distribution of spectral types for a star-forming region.
As shown in Figure~\ref{fig:histo}, peaks for both IC~348 and Chamaeleon~I
are rather narrow, which indicates that most of the low-mass members
are not experiencing large variations in spectral type ($>1$~subclass).
If that was the case, then the spectral type peak would be much broader,
and the abrupt decline later than M5 would not be present.
Furthermore, the possible presence of activity-induced errors in
spectral types does not fundamentally change the outlook for the IMF's
accuracy since comparable errors are likely already present.
For instance, for a typical member of a star-forming region, the
published spectral types often differ by a few subclasses, and
these classification errors can be either random or systematic
(the members of IC~348 and Chamaeleon~I in Figure~\ref{fig:histo} have been
classified in an unusually uniform manner).
Derivations of IMFs for young clusters are subject to various other sources
of error, such as the choice of temperature scale and the evolutionary models.
However, all measurements of the IMF, regardless of the type of stellar
population, are subject to random and systematic errors, many of which
are often difficult to quantify.
The relevant question is not whether such errors exist, but
whether the final IMF produced by a given methodology is accurate.
The random errors for masses in young clusters will simply
smooth out any small-scale structure in the IMF, and thus are unimportant
for the measurements of broad features and slopes, particularly given the
large sizes of the mass bins that are generally used.
Meanwhile, systematic errors in mass estimates will stretch or contract the
IMF, perhaps to a different degree as a function of mass. The agreement
between the IMFs of star-forming regions, open clusters, and the field
indicates that large errors of this kind are not present at masses
higher than $\sim0.02$~$M_{\odot}$ \citep[Figure~\ref{fig:imf},][]{luh06ab}.
However, IMF estimates in young clusters are untested at lower masses, and
could have significant errors given the peculiar photometric
properties of young L and T dwarfs and the resulting uncertainties in their
temperatures and luminosities.

\section{Concluding Remarks}

The absence of a significant dependence of the abundance of brown dwarfs 
on stellar density or the presence of O stars,
the ability of brown dwarfs to form in isolation and in wide binaries,
and the likely existence of protostellar brown dwarfs together provide
compelling evidence that brown dwarfs can form without the involvement of
tidal shear in massive cluster-forming cores, dynamical interactions, disk
fragmentation around solar-type stars, or photoionizing radiation.
Thus, it seems likely that the one remaining formation mechanism from
Section~\ref{sec:theory}, turbulent fragmentation, is responsible for some
fraction of brown dwarfs, and perhaps most of the ones in low-density regions
like Taurus. 
It is possible that other proposed mechanisms also produce brown dwarfs,
particularly in dense clusters, although there
is not any clear observational evidence of this so far.
One of the most promising avenues for better understanding the formation
of brown dwarfs is continued study of low-mass protostars by
identifying them in larger numbers with data from
{\it Spitzer}, WISE, and {\it Herschel Observatory} and by detailed
followup observations with facilities like the Atacama Large Millimeter Array.

Measurements of dynamical masses for a small number of young low-mass stars
and brown dwarfs have confirmed the theoretical prediction that
the hydrogen burning mass limit occurs near a spectral type of M6 for
ages of $\sim1$--100~Myr.  Based on a wide variety of studies,
young objects later than $\sim$M9, including planetary-mass companions,
exhibit unusually red colors and faint absolute magnitudes at near-IR
wavelengths relative to older field dwarfs.
It appears that this behavior can be explained with model atmospheres
that have clouds, low gravities, and non-equilibrium chemistry.
Further testing and refinement of the atmospheric and evolutionary models at 
young ages and low masses will require additional measurements of dynamical
masses and larger samples of well-characterized young L and T dwarfs. 

\section*{ACKNOWLEDGMENTS}

We thank John Bochanski, Jacqueline Faherty, Eric Feigelson, Lee Hartmann,
William Herbst, Adam Kraus, Dagny Looper, Eric Mamajek, and John Stauffer
for access to unpublished results and comments on the manuscript.
We acknowledge support from grant AST-0544588 from the National Science
Foundation. The Center for Exoplanets and Habitable Worlds is supported by the 
Pennsylvania State University, the Eberly College of Science, and the 
Pennsylvania Space Grant Consortium.

\bibliographystyle{Astronomy}
\bibliography{references}

\section{Sidebar}

\subsection{The Discovery of Brown Dwarfs}

The following should appear near the introduction:

The existence of brown dwarfs was predicted in the 1960's \citep{kum63,hay63},
but they were not found until more than two decades later.
One of the first promising candidates was GD 165~B, which was discovered
as a companion to a white dwarf \citep{bec88}.
This object was the first known member of the L spectral class of
cool dwarfs \citep[$T_{\rm eff}=1500$--2500~K,][]{kir93} and could be either a
very low-mass star or a brown dwarf \citep{kir99b}.
Through radial velocity measurements of the star HD~114762, \citet{lat89}
detected the presence of a close, unseen companion with a minimum mass
of 11~$M_{\rm Jup}$. It is likely a massive planet or a brown dwarf,
although its substellar nature is not guaranteed since the inclination of
its orbit is unknown.
The first unambiguous example of a brown dwarf was the 
companion Gl229~B \citep{nak95}, which exhibited strong methane absorption
that firmly established that it was too cool to be a star \citep{opp95}.
Near the time that Gl229~B was discovered, PPl~15 and Teide~1 were
identified as promising brown dwarf candidates in the Pleiades open cluster
\citep{stau94,reb95}. They were confirmed as substellar by the detection of Li
absorption \citep{bas96,reb96}, making them the first known free-floating
brown dwarfs.

\end{document}